\documentclass{iau} 
\usepackage{graphicx}
\usepackage{color}

\usepackage[T1]{fontenc}
\usepackage[applemac]{inputenc} 

\newcommand{\todo}{\ifmmode \text{\color{red}\Huge{\(\bullet\)}} \else {\color{red}{\Huge$\bullet$}}\fi}
\newcommand{\tido}{\ifmmode {{\color{red}\bullet}} \else {\color{red}$\bullet$}\fi}

\newcommand{\E        }[1]{\ifmmode 10^{#1} \else $10^{#1}$\fi}
\newcommand{\tE        }[1]{\ifmmode \times10^{#1} \else $\times10^{#1}$\fi}
\newcommand{\til}{\ifmmode \sim \else $\sim$\fi}

\newcommand{\et}{et al.\ }

\newcommand{\pc}	{\ifmmode {\rm pc} \else pc\fi}
\newcommand{\kpc}	{\ifmmode {\rm kpc} \else kpc\fi}
\newcommand{\ld}	{\ifmmode {\rm l.d.} \else l.d.\fi}
\newcommand{\kms}	{\ifmmode {\rm km\,s}^{-1} \else km\,s$^{-1}$\fi}
\newcommand{\cc}	{\ifmmode {\rm cm}^{-3}    \else cm$^{-3}$\fi}
\newcommand{\cmii}	{\ifmmode {\rm cm}^{-2}    \else cm$^{-2}$\fi}
\newcommand{\ergs}	{\ifmmode {\rm erg\,s}^{-1} \else erg s$^{-1}$\fi}
\newcommand{\ergcms}	{\ifmmode {\rm erg\,cm}^{-2}\,{\rm s}^{-1} \else erg\,cm$^{-2}$\,s$^{-1}$\fi}
\newcommand{\ergcmsA}	{\ifmmode {\rm erg\,cm}^{-2}\,{\rm s}^{-1}\,{\rm\AA}^{-1}
\else erg\,cm$^{-2}$\,s$^{-1}$\,\AA$^{-1}$\fi}
\newcommand{  \ergcmsHz  }{\ifmmode{\rm erg\,cm}^{-2}\,{\rm s}^{-1}\,{\rm Hz}^{-1}
                       \else ergs\,cm$^{-2}$\,s$^{-1}$\,Hz$^{-1}$\fi}
\newcommand{\kev}	{\ifmmode {\rm keV} \else keV\fi}

\newcommand{\mic}	{\ifmmode {\rm \mu m} \else $\mu$m\fi}
\newcommand{\vFWHM}	{\ifmmode v_{\mbox{\tiny FWHM}} \else $v_{\mbox{\tiny FWHM}}$\fi}
\newcommand{\vBLR}	{\ifmmode v_{\mbox{\tiny BLR}} \else $v_{\mbox{\tiny BLR}}$\fi}
\newcommand{\sigBLR}	{\ifmmode \sigma_{\mbox{\tiny BLR}} \else $\sigma_{\mbox{\tiny BLR}}$\fi}
\newcommand{\vNLR}	{\ifmmode v_{\mbox{\tiny NLR}} \else $v_{\mbox{\tiny NLR}}$\fi}
\newcommand{\tauBLR}	{\ifmmode \tau_{\mbox{\tiny BLR}} \else $\tau_{\mbox{\tiny BLR}}$\fi}

\newcommand{\Hubble}	{\ifmmode {\rm km\,s}^{-1}\,{\rm Mpc}^{-1} \else km\,s$^{-1}$\,Mpc$^{-1}$\fi}
\newcommand{\NDunit}	{\ifmmode {\rm Mpc}^{-3} \else Mpc$^{-3}$\fi}
\newcommand{\LFunit}	{\ifmmode {\rm Mpc}^{-3}\,{\rm mag}^{-1} \else Mpc$^{-3}$\,mag$^{-1}$\fi}
\newcommand{\MFunit}	{\ifmmode {\rm Mpc}^{-3}\,{\rm dex}^{-1} \else Mpc$^{-3}$\,dex$^{-1}$\fi}

\newcommand{\Msun}{\ifmmode M_{\odot} \else $M_{\odot}$\fi}
\newcommand{\Lsun}{\ifmmode L_{\odot} \else $L_{\odot}$\fi}
\newcommand{\Zsun}{\ifmmode Z_{\odot} \else $Z_{\odot}$\fi}
\newcommand{\mpyr}{\ifmmode \Msun\,{\rm yr}^{-1} \else $\Msun\,{\rm yr}^{-1}$\fi}

\newcommand{\Msol}{\Msun}

\newcommand{\qnote}{\ifmmode q_{0} \else $q_{0}$\fi}
\newcommand{\Hnote}{\ifmmode H_{0} \else $H_{0}$\fi}
\newcommand{\hnote}{\ifmmode h_{0} \else $h_{0}$\fi}
\newcommand{\anote}{\ifmmode a_{0} \else $a_{0}$\fi}
\newcommand{\tnote}{\ifmmode t_{0} \else $t_{0}$\fi}

\def\gsim{\;\rlap{\lower 2.5pt \hbox{$\sim$}}\raise 1.5pt\hbox{$>$}\;}
\def\lsim{\;\rlap{\lower 2.5pt \hbox{$\sim$}}\raise 1.5pt\hbox{$<$}\;}

\newcommand{  \Halpha   }{\ifmmode {\rm H}\alpha \else H$\alpha$\fi}

\newcommand{  \ha       }{\Halpha}
\newcommand{  \Hbeta    }{\ifmmode {\rm H}\beta \else H$\beta$\fi}

\newcommand{  \hb       }{\Hbeta}
\newcommand{  \Hgamma   }{\ifmmode {\rm H}\gamma \else H$\gamma$\fi}
\newcommand{  \Hdelta   }{\ifmmode {\rm H}\delta \else H$\delta$\fi}
\newcommand{  \Lya      }{\ifmmode {\rm Ly}\alpha \else Ly$\alpha$\fi}
\newcommand{  \Lyb      }{\ifmmode {\rm Ly}\beta \else Ly$\beta$\fi}
\newcommand{  \Pa       }{\ifmmode {\rm P}\alpha \else P$\alpha$\fi}
\newcommand{  \Pb       }{\ifmmode {\rm P}\beta \else P$\beta$\fi}
\newcommand{  \Bra      }{\ifmmode {\rm Br}\alpha \else Br$\alpha$\fi}
\newcommand{  \Brg      }{\ifmmode {\rm Br}\gamma \else Br$\gamma$\fi}
\newcommand{  \hii      }{\ifmmode {\rm H}\,\textsc{ii} \else H\,\textsc{ii}\fi}
\newcommand{  \hei      }{\ifmmode {\rm He}\,\textsc{i} \else He\,\textsc{i}\fi}
\newcommand{  \heii     }{\ifmmode {\rm He}\,\textsc{ii} \else He\,\textsc{ii}\fi}
\newcommand{  \HeIIuv   }{\ifmmode {\rm He}\,\textsc{ii}\,\lambda1640 \else He\,\textsc{ii}\,$\lambda1640$\fi}
\newcommand{  \HeIIop   }{\ifmmode {\rm He}\,\textsc{ii}\,\lambda4686 \else He\,\textsc{ii}\,$\lambda4686$\fi}
\newcommand{  \CII	}{\ifmmode \left[{\rm C}\,\textsc{ii}\right]\,\lambda157.74\,\mu{\rm m} \else [C\,{\sc ii}]\ $\lambda157.74\,\mu{\rm m}$\fi}
\newcommand{  \cii	}{\ifmmode \left[{\rm C}\,\textsc{ii}\right] \else [C\,{\sc ii}]\fi}

\newcommand{  \ciii     }{\ifmmode {\rm C}\,\textsc{iii}\right] \else C\,\textsc{iii}]\fi}
\newcommand{  \CIII     }{\ifmmode {\rm C}\,\textsc{iii}\right]\,\lambda1909 \else C\,\textsc{iii}]\,$\lambda1909$\fi}
\newcommand{  \civ      }{\ifmmode {\rm C}\,\textsc{iv}  \else C\,\textsc{iv}\fi}
\newcommand{  \CIV      }{\ifmmode {\rm C}\,\textsc{iv}\,\lambda1549 \else C\,\textsc{iv}\,$\lambda1549$\fi}
\newcommand{  \NIIopt   }{\ifmmode \left[{\rm N}\,\textsc{ii}\right]\,\lambda6584 \else [N\,\textsc{ii}]\,$\lambda6584$\fi}
\newcommand{  \nii      }{\ifmmode \left[{\rm N}\,\textsc{ii}\right]  \else [N\,\textsc{ii}]\fi}
\newcommand{  \niii     }{\ifmmode {\rm N}\,\textsc{iii} \else N\,\textsc{iii}\fi}
\newcommand{  \NIII     }{\ifmmode {\rm N}\,\textsc{iii}\,\lambda4640 \else N\,\textsc{iii}\,$\lambda4640$\fi}
\newcommand{  \niv      }{\ifmmode {\rm N}\,\textsc{iv}  \else N\,\textsc{iv}\fi}
\newcommand{  \NIVuv    }{\ifmmode {\rm N}\,\textsc{iv}\,\lambda1486 \else N\,\textsc{iv}\,$\lambda1486$\fi}
\newcommand{  \nv       }{\ifmmode {\rm N}\,\textsc{v}   \else N\,\textsc{v}\fi}
\newcommand{\oi}{\ifmmode \left[{\rm O}\,\textsc{i}\right] \else [O\,{\sc i}]\fi}
\newcommand{\OI}{\ifmmode \left[{\rm O}\,\textsc{i}\right]\,\lambda6300 \else [O\,{\sc i}]$\,\lambda6300$\fi}
\newcommand{\oii}{\ifmmode \left[{\rm O}\,\textsc{ii}\right] \else [O\,{\sc ii}]\fi}
\newcommand{\OII}{\ifmmode \left[{\rm O}\,\textsc{ii}\right]\,\lambda3727 \else [O\,{\sc ii}]\,$\lambda3727$\fi}
\newcommand{\oiii}{\ifmmode \left[{\rm O}\,\textsc{iii}\right] \else [O\,{\sc iii}]\fi}
\newcommand{\OIII}{\ifmmode \left[{\rm O}\,\textsc{iii}\right]\,\lambda5007 \else [O\,{\sc iii}]\,$\lambda5007$\fi}
\newcommand{  \OIIIbf   }{\ifmmode {\rm O}\,\textsc{iii}\,\lambda3133 \else O\,\textsc{iii}\,$\lambda3133$\fi}
\newcommand{  \OIIIuv   }{\ifmmode {\rm O}\,\textsc{iii}\,\lambda1663 \else O\,\textsc{iii}\,$\lambda1663$\fi}
\newcommand{  \oiv      }{\ifmmode {\rm O}\,\textsc{iv}  \else O\,\textsc{iv}\fi}
\newcommand{  \OIVuv    }{\ifmmode {\rm O}\,\textsc{iv}\,\lambda1402  \else O\,\textsc{iv}\,$\lambda1402$\fi}
\newcommand{  \OIVIR    }{\ifmmode {\rm O}\,\textsc{iv}\,25.9\,\mu {\rm m} \else O\,\textsc{iv}\,$25.9\,\mu$m\fi}
\newcommand{  \ovi      }{\ifmmode {\rm O}\,\textsc{vi}   \else O\,\textsc{vi}\fi}
\newcommand{  \Ovi      }{\ifmmode {\rm O}\,\textsc{vi}\,\lambda1035 \else O\,\textsc{vi}\,$\lambda1035$\fi}
\newcommand{  \nei      }{\ifmmode {\rm Ne}\,\textsc{i}   \else Ne\,\textsc{i}\fi}
\newcommand{  \neii     }{\ifmmode {\rm Ne}\,\textsc{ii}  \else Ne\,\textsc{ii}\fi}
\newcommand{  \NeiiIR   }{\ifmmode {\rm Ne}\,\textsc{ii}\,12.8\,\mu {\rm m} \else Ne\,\textsc{ii}\,$12.8\,\mu$m\fi}
\newcommand{  \neiii    }{\ifmmode {\rm Ne}\,\textsc{iii} \else Ne\,\textsc{iii}\fi}
\newcommand{  \neiv     }{\ifmmode {\rm Ne}\,\textsc{iv}  \else Ne\,\textsc{iv}\fi}
\newcommand{  \nev      }{\ifmmode {\rm Ne}\,\textsc{v}   \else Ne\,\textsc{v}\fi}
\newcommand{  \NevIR    }{\ifmmode {\rm Ne}\,\textsc{v}\,24.3\,\mu {\rm m} \else Ne\,\textsc{v}\,$24.3\,\mu$m\fi}
\newcommand{  \nevi     }{\ifmmode {\rm Ne}\,\textsc{vi}  \else Ne\,\textsc{vi}\fi}
\newcommand{  \mgi      }{\ifmmode {\rm Mg}\,\textsc{i} \else Mg\,\textsc{i}\fi}
\newcommand{  \mgii     }{\ifmmode {\rm Mg}\,\textsc{ii} \else Mg\,\textsc{ii}\fi}
\newcommand{  \MgII     }{\ifmmode {\rm Mg}\,\textsc{ii}\,\lambda2798 \else Mg\,\textsc{ii}\,$\lambda2798$\fi}
\newcommand{  \sii      }{\ifmmode {\rm S}\,\textsc{ii} \else S\,\textsc{ii}\fi}
\newcommand{  \siii     }{\ifmmode {\rm S}\,\textsc{iii} \else S\,\textsc{iii}\fi}
\newcommand{  \siv      }{\ifmmode {\rm S}\,\textsc{iv} \else S\,\textsc{iv}\fi}
\newcommand{  \sili     }{\ifmmode {\rm Si}\,\textsc{i}   \else Si\,\textsc{i}\fi}
\newcommand{  \silii    }{\ifmmode {\rm Si}\,\textsc{ii}  \else Si\,\textsc{ii}\fi}
\newcommand{  \Siliv    }{\ifmmode {\rm Si}\,\textsc{iv}  \else Si\,\textsc{iv}\fi}
\newcommand{  \SilIVuv  }{\ifmmode {\rm Si}\,\textsc{iv}\,\lambda1400  \else Si\,\textsc{iv}\,$\lambda1400$\fi}
\newcommand{  \AlIII   }{\ifmmode {\rm Al}\,\textsc{iii}\,\lambda1857 \else Al\,\textsc{iii}\,$\lambda1857$\fi}
\newcommand{  \Aliii   }{\ifmmode {\rm Al}\,\textsc{iii} \else Al\,\textsc{iii}\fi}
\newcommand{  \caii     }{\ifmmode {\rm Ca}\,\textsc{ii} \else Ca\,\textsc{ii}\fi}
\newcommand{  \feii     }{\ifmmode {\rm Fe}\,\textsc{ii} \else Fe\,\textsc{ii}\fi}
\newcommand{  \feiii    }{\ifmmode {\rm Fe}\,\textsc{iii} \else Fe\,\textsc{iii}\fi}
\newcommand{  \Kalpha   }{\ifmmode {\rm K}\alpha \else K$\alpha$\fi}

\newcommand{ \Lhb   }{\ifmmode L_{\hb} \else $L_{\hb}$\fi}
\newcommand{ \Lha   }{\ifmmode L_{\ha} \else $L_{\ha}$\fi}
\newcommand{ \fwhb  }{\ifmmode {\rm FWHM}\left(\hb\right) \else FWHM(\hb)\fi}
\newcommand{\sighb  }{\ifmmode \sigma\left(\hb\right) \else $\sigma\left(\hb\right)$\fi}
\newcommand{ \ewhb  }{\ifmmode {\rm EW}\left(\hb\right) \else EW(\hb)\fi}
\newcommand{ \fwha  }{\ifmmode {\rm FWHM}\left(\ha\right) \else FWHM(\ha)\fi}
\newcommand{ \ewha  }{\ifmmode {\rm EW}\left(\ha\right) \else EW(\ha)\fi}
\newcommand{ \Lmg   }{\ifmmode L\left(\mgii\right) \else $L\left(\mgii\right)$\fi}
\newcommand{ \fwmg  }{\ifmmode {\rm FWHM}\left(\mgii\right) \else FWHM(\mgii)\fi}
\newcommand{ \Lciv  }{\ifmmode L\left(\civ\right) \else $L\left(\civ\right)$\fi}
\newcommand{ \fwciv }{\ifmmode {\rm FWHM}\left(\civ\right) \else FWHM(\civ)\fi}
\newcommand{ \fwhm  }{\ifmmode {\rm FWHM} \else FWHM\fi} 
\newcommand{ \voff  }{\ifmmode v_{\rm off} \else $v_{\rm off}$\fi} 
\newcommand{ \vmax  }{\ifmmode v_{\rm max} \else $v_{\rm max}$\fi} 

\newcommand{ \mumg  }{\ifmmode \mu\left(\mgii\right) \else $\mu\left(\mgii\right)$\fi}
\newcommand{ \fmg   }{\ifmmode f\left(\mgii\right) \else $f\left(\mgii\right)$\fi}
\newcommand{ \muciv }{\ifmmode \mu\left(\civ\right) \else $\mu\left(\civ\right)$\fi}
\newcommand{ \fciv  }{\ifmmode f\left(\civ\right) \else $f\left(\civ\right)$\fi}

\newcommand{  \auvo     }{\ifmmode \alpha_{\nu,{\rm UVO}} \else $\alpha_{\nu,{\rm UVO}}$\fi}
\newcommand{  \Ledd     }{\ifmmode L_{\rm Edd} \else $L_{\rm Edd}$\fi}
\newcommand{  \lamLlam  }{\ifmmode \lambda L_{\lambda} \else $\lambda L_{\lambda}$\fi}
\newcommand{  \lLl      }{\ifmmode \lambda L_{\lambda} \else $\lambda L_{\lambda}$\fi}
\newcommand{  \nuLnu    }{\ifmmode \nu L_{\nu} \else $\nu L_{\nu}$\fi}
\newcommand{  \nLn      }{\ifmmode \nu L_{\nu} \else $\nu L_{\nu}$\fi}
\newcommand{  \Luv      }{\ifmmode L_{1450} \else $L_{1450}$\fi}
\newcommand{  \Lop      }{\ifmmode L_{5100} \else $L_{5100}$\fi}
\newcommand{  \lLop     }{\ifmmode \log\left(\Lop/\ergs\right) \else $\log\left(\Lop/\ergs\right)$\fi}
\newcommand{  \Lthree   }{\ifmmode L_{3000} \else $L_{3000}$\fi}
\newcommand{  \lLthree  }{\ifmmode \log\left(\Lthree/\ergs\right) \else $\log\left(\Lthree/\ergs\right)$\fi}
\newcommand{  \Lsix      }{\ifmmode L_{6200} \else $L_{6200}$\fi}
\newcommand{  \lLisx     }{\ifmmode \log\left(\Lop/\ergs\right) \else $\log\left(\Lop/\ergs\right)$\fi}
\newcommand{  \Lxray    }{\ifmmode L_{\rm X} \else $L_{\rm X}$\fi}

\newcommand{  \Lhard    }{\ifmmode L_{\rm 2-10} \else $L_{\rm 2-10}$\fi}
\newcommand{  \Lsoft    }{\ifmmode L_{\rm 0.5-2} \else $L_{\rm 0.5-2}$\fi}

\newcommand{\Fthree}{\ifmmode F_{3000} \else $F_{3000}$\fi}
\newcommand{\fuv}{\ifmmode f_{\lambda}\left(1450{\rm \AA}\right) \else $f_{\lambda}\left(1450 {\rm \AA}\right)$\fi}
\newcommand{\fthree}{\ifmmode f_{\lambda}\left(3000{\rm \AA}\right) \else $f_{\lambda}\left(3000{\rm \AA}\right)$\fi}
\newcommand{\fH}{\ifmmode f_{\lambda}\left(1.65\micron\right) \else
$f_{\lambda}\left(1.65\micron\right)$\fi}

\newcommand{\fbol}{\ifmmode f_{\rm bol} \else $f_{\rm bol}$\fi}
\newcommand{\fbolwv}{\ifmmode f_{\rm bol}\left(\lambda\right) \else $f_{\rm bol}\left(\lambda\right)$\fi}
\newcommand{\fbolopt}{\ifmmode f_{\rm bol}\left(5100{\rm \AA}\right) \else $f_{\rm bol}\left(5100{\rm \AA}\right)$\fi}
\newcommand{\fbolthree}{\ifmmode f_{\rm bol}\left(3000{\rm \AA}\right) \else $f_{\rm bol}\left(3000{\rm \AA}\right)$\fi}
\newcommand{\fboluv}{\ifmmode f_{\rm bol}\left(1450{\rm \AA}\right) \else $f_{\rm bol}\left(1450{\rm \AA}\right)$\fi}

\newcommand{\fbolbat}{\ifmmode f_{\rm bol}\left(14-150\,\kev\right) \else $f_{\rm bol}\left(14-150\,\kev\right)$\fi}
\newcommand{\fbolhard}{\ifmmode f_{\rm bol}\left(2-10\,\kev\right) \else $f_{\rm bol}\left(2-10\,\kev\right)$\fi}

\newcommand{\fobs}{\ifmmode f_{\rm obs} \else $f_{\rm obs}$\fi}

\newcommand{  \mbh      }{\ifmmode M_{\rm BH} \else $M_{\rm BH}$\fi}
\newcommand{  \lmbh     }{\ifmmode \log\left(\mbh/\Msun\right) \else $\log\left(\mbh/\Msun\right)$\fi} 
\newcommand{  \lledd    }{\ifmmode L/L_{\rm Edd} \else $L/L_{\rm Edd}$\fi}
\newcommand{  \mmedd    }{\ifmmode \dot{m}/\dot{m}_{\rm \,Edd} \else $\dot{m}/\dot{m}_{\rm \,Edd}$\fi}
\newcommand{  \Lbol     }{\ifmmode L_{\rm bol} \else $L_{\rm bol}$\fi}
\newcommand{  \lbol     }{\ifmmode L_{\rm bol} \else $L_{\rm bol}$\fi}
\newcommand{  \lLbol    }{\ifmmode \log\left(\Lbol/\ergs\right) \else $\log\left(\Lbol/\ergs\right)$\fi} 
\newcommand{  \Lagn     }{\ifmmode L_{\rm AGN} \else $L_{\rm AGN}$\fi}
\newcommand{  \lagn     }{\ifmmode L_{\rm AGN} \else $L_{\rm AGN}$\fi}

\newcommand{  \tgrow     }{\ifmmode t_{\rm growth} \else $t_{\rm growth}$\fi}
\newcommand{  \tAD     }{\ifmmode t_{\rm acc} \else $t_{\rm acc}$\fi}
\newcommand{  \tacc    }{\ifmmode t_{\rm acc} \else $t_{\rm acc}$\fi}
\newcommand{  \tUni      }{\ifmmode t_{\rm Universe} \else $t_{\rm Universe}$\fi}

\newcommand{  \Mdotin	}{\ifmmode \dot{M}_{\rm infall} \else $\dot{M}_{\rm infall}$\fi}
\newcommand{  \Mdotbh	}{\ifmmode \dot{M}_{\rm BH} \else $\dot{M}_{\rm BH}$\fi}
\newcommand{  \Mdotad	}{\ifmmode \dot{M}_{\rm AD} \else $\dot{M}_{\rm AD}$\fi}
\newcommand{  \Mdotacc	}{\ifmmode \dot{M}_{\rm acc} \else $\dot{M}_{\rm acc}$\fi}
\newcommand{  \Mdotthin	}{\ifmmode \dot{M}_{\rm thin} \else $\dot{M}_{\rm thin}$\fi}
\newcommand{  \Mdotdisk	}{\ifmmode \dot{M}_{\rm disk} \else $\dot{M}_{\rm disk}$\fi}

\newcommand{  \Mindot	}{\ifmmode \dot{M}_{\rm infall} \else $\dot{M}_{\rm infall}$\fi}
\newcommand{  \Mbhdot	}{\ifmmode \dot{M}_{\rm BH} \else $\dot{M}_{\rm BH}$\fi}
\newcommand{  \Maddot	}{\ifmmode \dot{M}_{\rm AD} \else $\dot{M}_{\rm AD}$\fi}
\newcommand{  \Maccdot	}{\ifmmode \dot{M}_{\rm acc} \else $\dot{M}_{\rm acc}$\fi}
\newcommand{  \Mthdot	}{\ifmmode \dot{M}_{\rm thin} \else $\dot{M}_{\rm thin}$\fi}
\newcommand{  \Mdsdot	}{\ifmmode \dot{M}_{\rm disk} \else $\dot{M}_{\rm disk}$\fi}

\newcommand{  \aox    }{\ifmmode \alpha_{\rm ox} \else $\alpha_{\rm ox}$\fi}

\newcommand{  \as	}{\ifmmode a_{\rm *} \else $a_{\rm *}$\fi}
\newcommand{  \avec	}{\ifmmode \vec{a}_{\rm *} \else $\vec{a}_{\rm *}$\fi}
\newcommand{  \re	}{\ifmmode \eta      	 \else $\eta$\fi}
\newcommand{  \RISCO	}{\ifmmode R_{\rm ISCO}  \else $R_{\rm ISCO}$\fi}

\newcommand{  \mseed    }{\ifmmode M_{\rm seed} \else $M_{\rm seed}$\fi}
\newcommand{  \mbul     }{\ifmmode M_{\rm bulge} \else $M_{\rm bulge}$\fi} 
\newcommand{  \mstar    }{\ifmmode M_{*} \else $M_{*}$\fi} 
\newcommand{  \mgal     }{\ifmmode M_{*} \else $M_{*}$\fi} 
\newcommand{  \mhost    }{\ifmmode M_{\rm host} \else $M_{\rm host}$\fi}
\newcommand{  \mmsmall  }{\ifmmode M_{\rm BH}/M_{*} \else $M_{\rm BH}/M_{*}$\fi}
\newcommand{  \mmlarge  }{\ifmmode M_{*}/M_{\rm BH} \else $M_{*}/M_{\rm BH}$\fi}

\newcommand{  \mmdotlarge}{\ifmmode \dot{M}_*/\Mbhdot \else $\dot{M}_*/\Mbhdot$\fi}
\newcommand{  \mmdotsmall}{\ifmmode \Mbhdot/\dot{M}_* \else $\Mbhdot/\dot{M}_*$\fi}

\newcommand{  \mmwp     }{\ifmmode \left(M_{*}/M_{\rm BH}\right) \else $\left(M_{*}/M_{\rm BH}\right)$\fi}
\newcommand{  \ml       }{\ifmmode M_{*}/L_{*} \else $M_{*}/L_{*}$\fi}
\newcommand{  \mlwp     }{\ifmmode \left(M_{*}/L\right) \else $\left(M_{*}/L\right)$\fi}
\newcommand{  \mlk      }{\ifmmode \left(M_{*}/L_{K}\right) \else $\left(M_{*}/L_{K}\right)$\fi}
\newcommand{  \sigs     }{\ifmmode \sigma_{*} \else $\sigma_{*}$\fi}
\newcommand{  \Reff     }{\ifmmode R_{\rm e} \else $R_{\rm e}$\fi}
\newcommand{  \Rvir     }{\ifmmode R_{\rm vir} \else $R_{\rm vir}$\fi}
\newcommand{  \Rtwo     }{\ifmmode R_{200} \else $R_{200}$\fi}
\newcommand{  \Rfive    }{\ifmmode R_{500} \else $R_{500}$\fi}
\newcommand{  \Rgrp     }{\ifmmode R_{\rm grp} \else $R_{\rm grp}$\fi}
\newcommand{  \nser     }{\ifmmode n_{\rm s} \else $n_{\rm s}$\fi}
\newcommand{  \LSF      }{\ifmmode L_{\rm SF}  \else $L_{\rm SF}$\fi}
\newcommand{  \LFIR     }{\ifmmode L_{\rm FIR} \else $L_{\rm FIR}$\fi}
\newcommand{  \Lfir     }{\ifmmode L_{\rm FIR} \else $L_{\rm FIR}$\fi}
\newcommand{  \LTIR     }{\ifmmode L_{\rm TIR} \else $L_{\rm TIR}$\fi}
\newcommand{  \Ltir     }{\ifmmode L_{\rm TIR} \else $L_{\rm TIR}$\fi}

\newcommand{  \mdyn     }{\ifmmode M_{\rm dyn} \else $M_{\rm dyn}$\fi} 
\newcommand{  \mgas     }{\ifmmode M_{\rm gas} \else $M_{\rm gas}$\fi} 
\newcommand{  \sfr      }{\ifmmode {\rm SFR} \else SFR\fi}

\newcommand{ \Lcii     }{\ifmmode L_{\cii} \else $L_{\cii}$\fi}
\newcommand{ \fwcii  }{\ifmmode {\rm FWHM}\cii \else FWHM\cii\fi}

\newcommand{  \hst     }  {{\it HST}}
\newcommand{  \jwst    }  {{\it JWST}}

\newcommand{  \spitzer }  {{\it Spitzer}}

\newcommand{\bj}{\ifmmode b_{\rm J} \else $b_{\rm J}$\fi}

\newcommand{\iab}{\ifmmode i_{\rm AB} \else $i_{\rm AB}$\fi}

\newcommand{\jab}{\ifmmode J_{\rm AB} \else $J_{\rm AB}$\fi}
\newcommand{\hab}{\ifmmode H_{\rm AB} \else $H_{\rm AB}$\fi}
\newcommand{\kab}{\ifmmode K_{\rm AB} \else $K_{\rm AB}$\fi}

\newcommand{\jveg}{\ifmmode J_{\rm Vega} \else $J_{\rm Vega}$\fi}
\newcommand{\hveg}{\ifmmode H_{\rm Vega} \else $H_{\rm Vega}$\fi}
\newcommand{\kveg}{\ifmmode K_{\rm Vega} \else $K_{\rm Vega}$\fi}

\newcommand{  \Chisq    }{\ifmmode \chi^{2} \else $\chi^{2}$}
\newcommand{  \nelec    }{\ifmmode n_{e} \else $n_{e}$\fi}     
\newcommand{  \nh       }{\ifmmode n_{\rm H} \else $n_{\rm H}$\fi}     
\newcommand{  \Ncol     }{\ifmmode N_{\rm col} \else $N_{\rm col}$\fi} 
\newcommand{  \NH       }{\ifmmode N_{\rm H} \else $N_{\rm H}$\fi}     

\def\ion#1#2{#1$\;${\small\rm\@Roman{#2}}\relax}

\title[The highest-$z$ quasars] 
{What do observations tell us about the highest-redshift supermassive black holes?}

\author[Benny Trakhtenbrot]   
{Benny Trakhtenbrot}

\affiliation{School of Physics and Astronomy, Tel Aviv University,\\ Tel Aviv 69978, Israel \\ email: {\tt benny@astro.tau.ac.il}}

\pubyear{2019}
\volume{356}  
\jname{Nuclear Activity in Galaxies Across Cosmic Times}
\editors{M. Povic et al., eds.}
\begin{document}

\maketitle

\begin{abstract}
I review the current understanding of some key properties of the earliest growing supermassive black holes (SMBHs), as determined from the most up-to-date observations of $z\gtrsim5$ quasars. 
This includes their accretion rates and growth history, their host galaxies, and the large-scale environments that enabled their emergence less than a billion years after the Big Bang. 
The available multi-wavelength data show that these SMBHs are consistent with Eddington-limited, radiatively efficient accretion that had to proceed almost continuously since very early epochs. 
ALMA observations of the hosts' ISM reveal gas-rich, well developed galaxies, with a wide range of SFRs that may exceed $\sim$1000 \mpyr. 
Moreover, ALMA uncovers a high fraction of companion, interacting galaxies, separated by $<$100 kpc (projected). 
This supports the idea that the first generation of high-mass, luminous SMBHs grew in over-dense environments, and that major mergers may be important drivers for rapid SMBH and host galaxy growth. 
Current X-ray surveys cannot access the lower-mass, supposedly more abundant counterparts of these rare $z\gtrsim5$ massive quasars, which should be able to elucidate the earliest stages of BH formation and growth. Such lower-mass nuclear BHs will be the prime targets of the deepest surveys planned for the next generation of facilities, such as the upcoming {\it Athena} mission and the future {\it Lynx} mission concept.
\keywords{quasars: general, black hole physics, galaxies: active, galaxies: high-redshift, galaxies: interactions}
\end{abstract}

\firstsection 

\section{Introduction and Overview}
\label{sec:intro}

For decades, quasars have been detected and studied at increasingly high redshifts, owing to their extremely high luminosities. 
Wide-area surveys, required by the rarity of luminous, high-redshift quasars, provided an almost continuous progress of breaking redshift records.  
Since the first few $z\gtrsim6$ quasars were detected through (early) SDSS observations (e.g., \cite[Fan et~al. 2003]{Fan2003_z6}), the trickle has turned into a steady stream of detections.
The current redshift record holder is the quasar ULAS J1342$+$0928 at $z{=}7.54$ (\cite[Ba{\~n}ados \et 2018a]{Banados2018_z75_Nature}), and there are over 170 quasars known at $z\gtrsim6$, in addition to over 300 at $z\sim5-6$ (see \cite[Ross \& Cross 2019]{Ross2019_z6_phot_cat}). 
The vast majority of these systems have been selected in various wide-field, multi-band optical-IR surveys, through elaborate colour-based criteria. 
Indeed, essentially every imaging survey with sufficient area and depth has identified samples of $z\gtrsim5$ quasars (for some of the largest relevant samples see, e.g., 
\cite[Willott \et 2010a]{Willott2010_QLF}, 
\cite[Ba{\~n}ados \et 2016]{Banados2016_PS1_z6_qsos}, 
\cite[Jiang \et 2016]{Jiang2016_SDSS_z6_final}, 
\cite[Reed \et 2017]{Reed2017_z6_QSOs_DES_VISTA}, 
\cite[Matsuoka \et 2019]{Matsuoka2019_SHELLQs_X_z6}, 
\cite[Yang \et 2018]{Yang2018_z55_survey}, 
\cite[Wang \et 2019a]{Wang2019_z65_QLF}). 
The quasar selection criteria are constantly improving, allowing to recover highly complete (or, at least, well-understood) samples that cover an ever expanding range in flux, redshift, and/or colour (e.g., \cite[Carnall \et 2015]{Carnall2015_z6_qsos}, \cite[Wang \et 2016]{Wang2016_WISE_z5_qsos}, \cite[Reed \et 2017]{Reed2017_z6_QSOs_DES_VISTA}). 
The recent, publicly accessible compilation of \cite[Ross \& Cross (2019)]{Ross2019_z6_phot_cat} provides an impressive and up-to-date status report on this still-growing population of quasars, as well as references to some of the important follow-up observations.
Most importantly, several teams have been accumulating a rich collection of multi-wavelength data for these systems, which allow to study a multitude of phenomena related to the quasars and their central engines, to their host galaxies, and indeed to their large-scale environments.

A key motivation for studying the nature of the highest-redshift quasars, their hosts and their environments, is the very existence of such extreme systems, which are powered by super-massive black holes (SMBHs) with masses of $\mbh\gtrsim10^8\,\Msol$. 
How could they grow to such high masses in less than a Gyr after the Big Bang (or seed BH formation)?
Could they have grown through extremely fast, super-Eddington accretion?
What do their host galaxies tell us about the availability of the (cold) gas that is needed for this fast, early growth?
What sort of large-scale environments are needed to form the earliest nuclear, massive BHs and to power their fast growth?
What are the effects that this fast SMBH growth exerts back on the host galaxies and/or the larger-scale environments, through radiative and/or mechanical energy output?

In this contribution, I review some of the main results related to the highest-redshift quasars, as inferred from {\it observations} across the electromagnetic spectrum, and over a wide range of physical scales.
I also discuss the whereabouts of the lower-luminosity, lower-mass counterparts of these quasars, and outline how upcoming and future facilities and surveys will further extend our understanding of the first generation of SMBHs.
For obvious reasons, I cannot fairly present {\it all} the recent results on this exciting topic.
Detailed discussions of some of the theoretical aspects relevant to this topic can be found in several reviews including, among others, those by \cite{Volonteri2010_rev}, \cite{Natarajan2011_seeds_rev}, \cite{Valiante2017_seeds_rev}, and \cite{Inayoshi2020_seeds_rev}.
In addition, \cite{Fan2006_reion_rev} reviews the usage of high-redshift quasars for probing the (re-)ionization state of their cosmic environments (and indeed of the  Universe) -- a topic that is not discussed here.

\section{Observed Properties of the Highest-Redshift Quasars}

\subsection{Basic spectral properties}

\begin{figure*}
    \centering
    \includegraphics[width=0.9\textwidth]
    {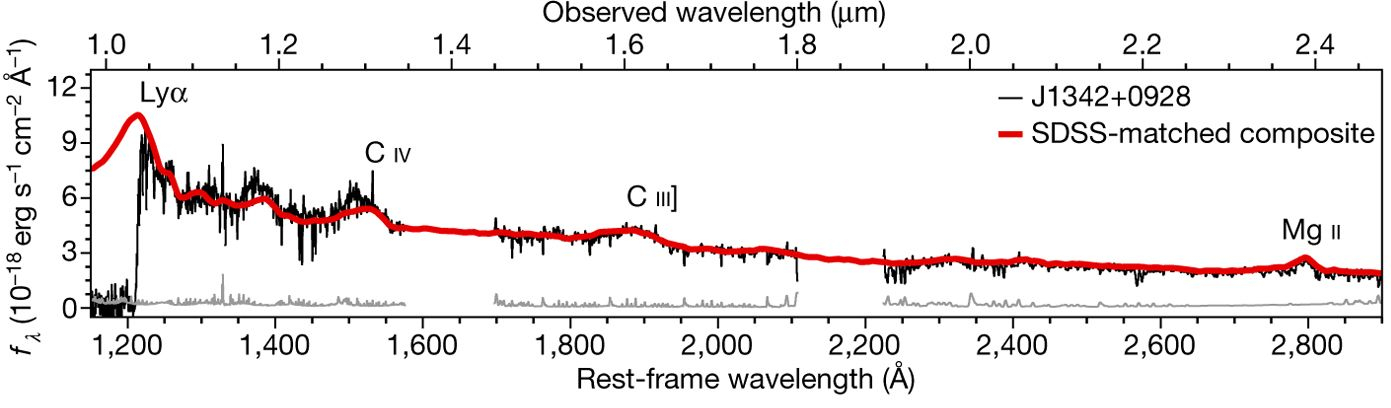}\\
    \vspace*{0.2cm}
    \includegraphics[width=0.9\textwidth]
    {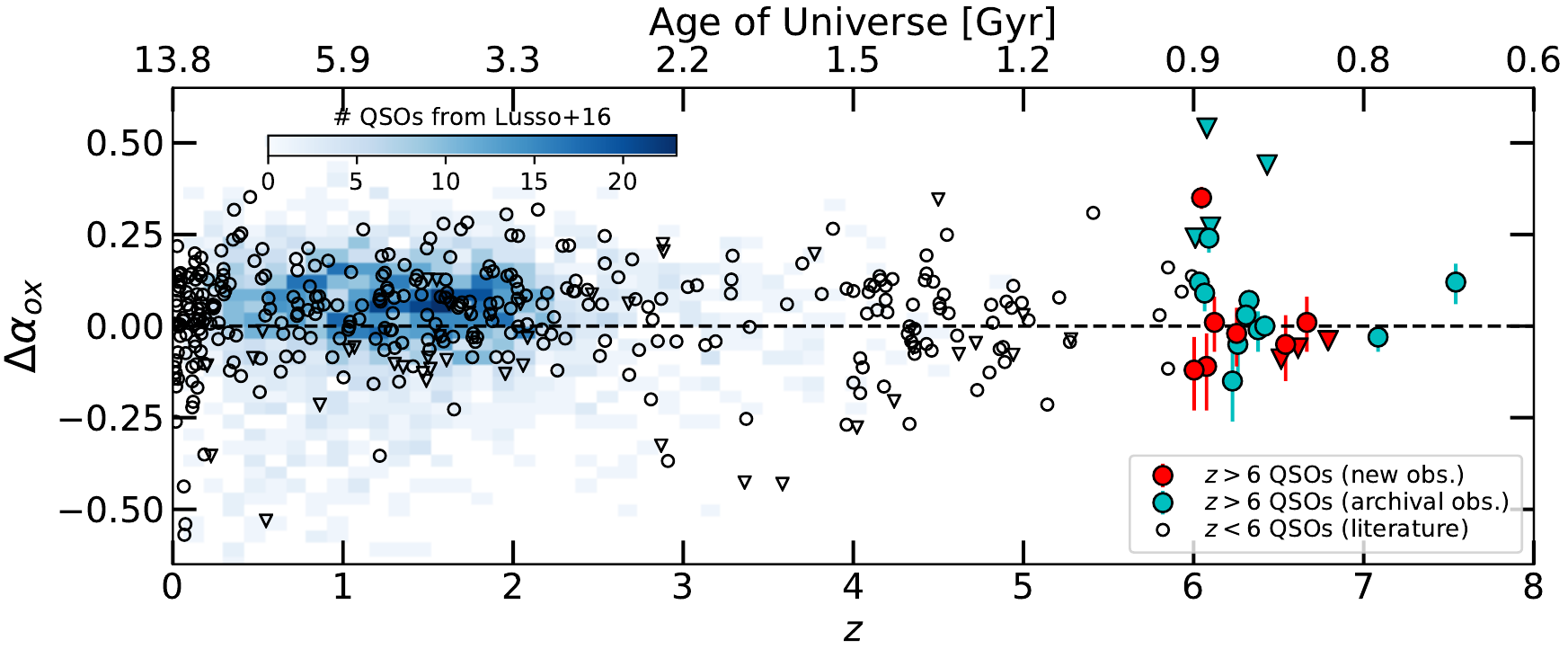}
    \caption{High-redshift quasars show ``normal'' UV-optical continuum and line emission, and broad-band (UV-to-X-ray) SEDs. 
    {\it Top:} the (rest-frame) near-UV spectrum of ULAS J1342+0928, the highest redshift quasar known to date ($z{=}7.54$), compared with a composite spectrum of lower-$z$ SDSS quasars (taken from \cite[Ba{\~n}ados \et 2018a]{Banados2018_z75_Nature}). 
    {\it Bottom:} the deviation in the optical-to-X-ray spectral slope, $\alpha_{\rm ox}$, compared with the $\aox-L_{\rm UV}$ relation calibrated at lower $z$ AGN (e.g., \cite[Lusso \& Risaliti 2016]{Lusso2016_Lx_Luv}), vs. redshift (figure taken from \cite[Vito \et 2019]{Vito2019_z6_Xrays}). 
    }
    \label{fig:spec_and_sed}
\end{figure*}

The first and perhaps most striking observation related to the highest-redshift quasars we know, is how ``normal'' their basic emission properties appear to be.

The (rest-frame) UV spectra of the highest-redshift quasars are remarkably similar to those of lower-redshift ones (matched in luminosity), including in particular the broad emission lines of \CIV, \CIII, and \MgII\ (Fig.~\ref{fig:spec_and_sed}, top).
Such comparisons have to account for the tendency of highly accreting quasars to show blue-shifted (UV) broad lines (e.g., \cite[Shen \et 2016]{Shen2016_SDSS_RM_shifts}, \cite[Mart{\'\i}nez-Aldama \et 2018]{MartinezAldama2018}, and references therein).
The relative intensities of these lines do not exhibit significant evolution out to $z\sim7$, suggestive of early metal enrichment in the dense circumnuclear gas that constitutes the broad line region (BLR; e.g., \cite[Jiang \et 2007]{Jiang2007}, \cite[Kurk \et 2007]{Kurk2007}, \cite[De Rosa \et 2011]{DeRosa2011}, \cite[2014]{DeRosa2014}, \cite[Mazzucchelli \et 2017]{Mazzucchelli2017_z65_props}).
The same can be said about the presence of hot, circumnuclear dust, which was detected through \spitzer\ mid-IR observations in many systems (\cite[Jiang \et 2006]{Jiang2006_Spitzer}), although in this case there is evidence that some systems are ``hot dust poor'' (\cite[Jiang \et 2010]{Jiang2010_HDP}).

An increasing number of $z\gtrsim6$ quasars have now been detected in the X-rays (e.g., \cite[Shemmer \et 2006]{Shemmer2006_CXO_hiz}, \cite[Ba{\~n}ados \et 2018b]{Banados2018_z75_Xray}, \cite[Vito \et 2019]{Vito2019_z6_Xrays}, \cite[Pons \et 2020]{Pons2020_Xray_z65}; see compilation by \cite[Nanni \et 2017]{Nanni2017}).
The strength of their X-ray emission follows the expectations from lower-redshift quasars, namely the close relation between the (rest-frame) UV luminosity and UV-to-X-ray spectral slope (\aox; e.g., \cite[Lusso \& Risaliti 2016]{Lusso2016_Lx_Luv}; Fig.~\ref{fig:spec_and_sed}, bottom).

A few of the highest-redshift quasars have been also detected in radio bands, implying extremely powerful radio emission, thought to originate from a relativistic jet (e.g., \cite[Ba{\~n}ados \et 2018c]{Banados2018_radio_z6}, \cite[Belladitta \et 2019]{Belladitta2019_z5_blazar_DES}).
If (some of) these systems are interpreted as beamed AGN (i.e., blazars; e.g., \cite[Sbarrato \et 2012]{Sbarrato2012_blazar_z53}, \cite[Ghisellini \et 2015]{Ghisellini2015_blazar_z52}), they may potentially allow us to further constrain the accretion efficiencies and space densities of {\it all} (actively) accreting SMBHs at $z\gtrsim5$ -- including those which are too faint to be picked up by current optical/NIR surveys (\cite[Volonteri \et 2011]{Volonteri2011_blazars}, \cite[Ghisellini \et 2013]{Ghisellini2013_blazars}).

\subsection{Mass and accretion rates}
\label{subsec:mass_and_acc}

\begin{figure*}
    \centering
    \includegraphics[trim={0.5cm, 1.0cm, 1.0cm, 0cm},clip,width=1.00\textwidth]
    {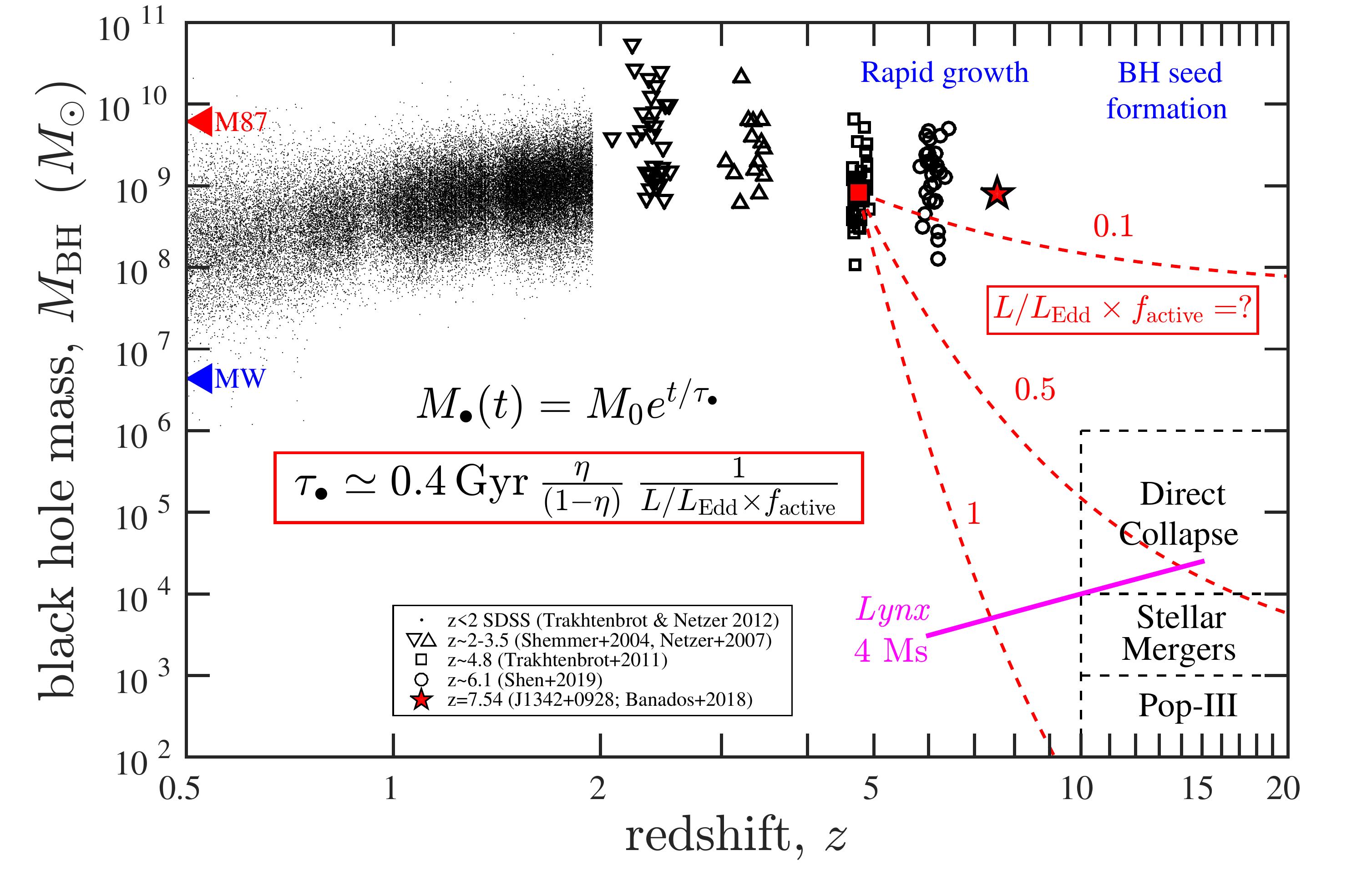}
    \caption{An overview of the emergence of the first generation of SMBHs. 
    The plot shows various samples of highly luminous quasars with reliable \mbh\ estimates (black symbols, see legend; only a subset of available measurements is shown). 
    Given the range of possible BH seed masses available at $z\sim10-20$ (black-dashed boxes), the $z\gtrsim5$ quasars require nearly-continuous mass accretion, at high accretion rates -- as illustrated by growth tracks with various combinations of \lledd\ and duty cycle (red-dashed lines).
    Future facilities, and particularly the {\it Lynx} mission concept, may be able to directly probe the epoch of massive seed BH formation and fast, initial BH growth -- the magenta line illustrates the expected sensitivity of a deep {\it Lynx} survey.}
    \label{fig:MBH_vs_z}
\end{figure*}

The high luminosities of the first quasars to be identified at $z\gtrsim5$, of $\Lbol\gtrsim10^{47}\,\ergs$, coupled with the Eddington argument (i.e., $\Lbol \lesssim L_{\rm Edd} \simeq 1.5{\times}10^{38}\,\left[\mbh/\Msol\right]\,\ergs$), immediately indicate that these sources are powered by SMBHs with masses of at least $\mbh\gtrsim10^9\,\Msol$.

More accurate estimates of \mbh\ were obtained by using dedicated, intensive near-IR spectroscopy, which probes the rest-frame UV broad \mgii\ emission line, and by relying on so-called ``virial'' (or ``single-epoch'') \mbh\ prescriptions (e.g., \cite[Trakhtenbrot \& Netzer 2012]{TrakhtNetzer2012_Mg2}, \cite[Shen 2013]{Shen2013_rev}).
These reveal that the highest redshift quasars are indeed powered by SMBHs with $\mbh\sim10^{8-10}\,\Msol$, accreting at rates $\lledd\sim0.1-1$ (e.g., 
\cite[Willott \et 2010b]{Willott2010_MBH},
\cite[Trakhtenbrot \et 2011]{Trakhtenbrot2011}, 
\cite[De Rosa \et 2014]{DeRosa2014}, 
\cite[Onoue \et 2019]{Onoue2019_SHELLQs_z6_MBH}, 
\cite[Shen \et 2019]{Shen2019_z6_NIR_spec}; see \cite[Wang \et 2015]{Wang2015_z5_hiM}, \cite[Wu \et 2015]{Wu2015_z6_nature}, and \cite[Kim \et 2018]{Kim2018_z6_low_LLEdd} for examples of ``extreme'' values).
Here, too, the number of sources with reliable determinations of \mbh\ and \lledd\ is steadily growing. 
For example, the recent study by \cite{Shen2019_z6_NIR_spec} reported new measurements of this sort for about 30 $z\gtrsim5.7$ quasars.

With \Lbol\ (and \lledd) estimates in hand, one still has to assume a certain radiative efficiency ($\eta{\equiv}\Lbol/\dot{M} c^2$) in order to deduce the physical accretion rates, and thus to address the dramatic growth history, of the earliest SMBHs.
A few studies used simple thin-disk models to deduce $\dot{M}$ (e.g., \cite[Sbarrato \et 2012]{Sbarrato2012_blazar_z53}, \cite[Trakhtenbrot \et 2017a]{Trakhtenbrot2017_z6_Mdot_eta}), showing that the (rest-frame) UV-optical SED measurements in hand can be indeed explained by Eddington-limited, radiatively efficient, thin-disk accretion. 
However, yet earlier episodes of super-Eddington accretion cannot be ruled out.

Comparing the known, luminous $z\gtrsim5$ quasars to the most luminous, highest-\mbh\ quasars at $z\sim2-4$, we finally see a {\it decrease} in the maximal \mbh\ with increasing redshift (Fig.~\ref{fig:MBH_vs_z}; see also \cite[Marziani \& Sulentic 2012]{MarzianiSulentic2012_MBH_rev} and \cite[Trakhtenbrot \& Netzer 2012]{TrakhtNetzer2012_Mg2}). 
The increase in typical \lledd\ means that we are witnessing the earliest epoch during which the most massive BHs known ($\mbh\sim10^{10}\,\Msol$) have been growing at their fastest, Eddington-limited rate (see discussion in, e.g., \cite[Trakhtenbrot \et 2011]{Trakhtenbrot2011}).

Most importantly, given the measured masses and accretion rates (and thin-disk values for $\eta$), the main challenge concerning the highest-redshift quasars still stands:  
these SMBHs had to grow nearly continuously, at high accretion rates, to reach their observed masses -- even if one allows for massive BH seeds.
This is illustrated in Fig.~\ref{fig:MBH_vs_z} using several simple, exponential mass growth tracks (i.e., constant \lledd), scaled to match a typical $z{\simeq}5$ quasar. 
Even the most massive BH seed ($M_{\rm seed}\sim10^6\,\Msol$; e.g., \cite[Volonteri 2010]{Volonteri2010_rev}, \cite[Natarajan 2011]{Natarajan2011_seeds_rev}) {\it cannot} explain the observed mass, if the growth proceeded through a combination of Eddington ratio and duty cycle of order $\lledd\times f_{\rm active} = 0.1$.

\subsection{Host galaxies}
\label{subsec:hosts}

Measuring the host galaxies of the earliest quasars is key to understanding their early emergence and fast growth, and to testing for any early signs of the links we may expect between SMBH and host growth (i.e., their co-evolution).
In particular, one would naively expect the earliest SMBHs to grow in unstable, gas-rich systems, which would also allow for intense SF. 
Major galaxy mergers, which are thought to be common at early cosmic epochs, could expedite SMBH and host growth. 
There are even some suggestions that the most massive, rarest luminous quasars, may have out-grown (the stellar populations of) their hosts, thus predicting the SMBHs to be ``over-massive'', compared with the locally observed $\mbh-\mhost$ relations (see, e.g., \cite[Agarwal \et 2013]{Agarwal2013_obese}).

\smallskip
All these ideas are extremely hard to address with observations. 
Indeed, to date there is no direct detection of the stellar component in the hosts of the highest-redshift quasars -- not even with intense, high-resolution \hst\ NIR imaging.
ALMA and other sub-mm facilities are currently revolutionising our understanding of the host galaxies, probing directly the (cold) ISM gas, and allowing us to determine the hosts' gas content, SF activity, and dynamics. 
Moreover, after a few earlier efforts focusing on particular systems, and thanks to the high sensitivity of ALMA, we are now seeing a shift towards spatially-resolved studies of increasingly large and systematically defined samples.

\smallskip
The vast majority of $z\gtrsim5$ quasar hosts are robustly detected, and often spatially resolved, in (rest-frame) FIR continuum emission. 
This implies considerable amounts of cold ($\sim$30-60 K) gas on galaxy-wide scales ($\gtrsim$1 kpc), and indeed that the highest redshift quasars are mostly hosted in well-developed galaxies, enriched in metals and dust.

\smallskip
The quasar hosts show a wide range of SFRs, reaching extremely high values. 
The higher-SFR systems, with ${\sim}1000-3000\,\mpyr$, were already clearly detected with {\it Herschel} (e.g., \cite[Mor \et 2012]{Mor2012_z48}, \cite[Leipski \et 2013]{Leipski2014_z6_FIR_SEDs}, \cite[Netzer \et 2014]{Netzer2014_z48_SFR}, \cite[Lyu \et 2016]{Lyu2016_FIR_SED}).
ALMA now allows us to probe much lower SFRs.
For example, \cite{Decarli2018_CII_z6} studied a sample of 27 quasars at $z\gtrsim6$, and found that their (rest-frame) FIR emission can be accounted for with $\sfr\sim30-2000\,\mpyr$.
Other ALMA-based studies report an even wider distribution, with SFRs as low as $\sim$10\,\mpyr\ (\cite[Willott \et 2017]{Willott2017_z6_SFR}, \cite[Izumi \et 2018]{Izumi2018_SHELLQs_III_z6_ALMA}), or as high as $\gtrsim$3000\,\mpyr\ (e.g., \cite[Trakhtenbrot \et 2017b]{Trakhtenbrot2017_z48_ALMA}).
The extremely high SFRs found for some $z\gtrsim5$ quasar hosts are consistent with the highest values known, typically measured in dusty, SF galaxies (i.e., ``sub-mm galaxies'', or SMGs; see \cite[Casey \et 2014]{Casey2014_SMGs_rev} for a review).
This is in line with the expectation for rapid, co-evolutionary growth of the BH and stellar components.
On the other hand, the lower end of the SFR distribution overlaps with what is known about the much more abundant, ``normal'' SF galaxies at $z\sim5-7$ (i.e., ``Lyman break galaxies'' or LBGs; see, e.g., \cite[Stark 2016]{Stark2016_hiz_gals_rev}).
Hopefully, we will soon be able to robustly test for possible links between the host SFR and SMBH-related properties, namely \Lagn, using complete samples -- as is done at lower redshifts (see initial attempts to populate the SFR-\Lagn\ parameter space in, e.g., \cite[Netzer \et 2014]{Netzer2014_z48_SFR}, \cite[Lyu \et 2016]{Lyu2016_FIR_SED}, \cite[Venemans \et 2018]{Venemans2018_z6_ALMA_cont}, and \cite[Izumi \et 2019]{Izumi2019_SHELLQs_VIII_z6_ALMA}).

\begin{figure*}
    \centering
    \includegraphics[trim={0 -7cm 0 -0.5cm},clip,width=0.3\textwidth]
    {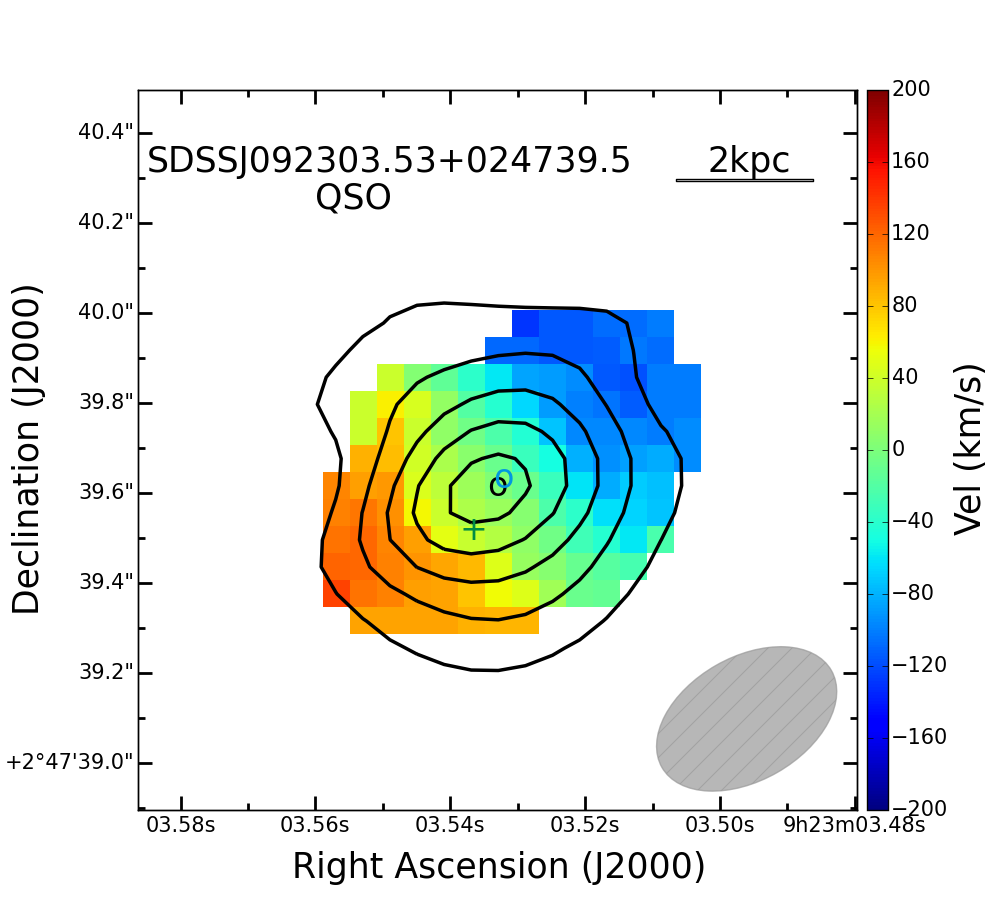}
    \includegraphics[trim={0.25cm 0.5cm 0.75cm 2.5cm},clip,width=0.675\textwidth]
    {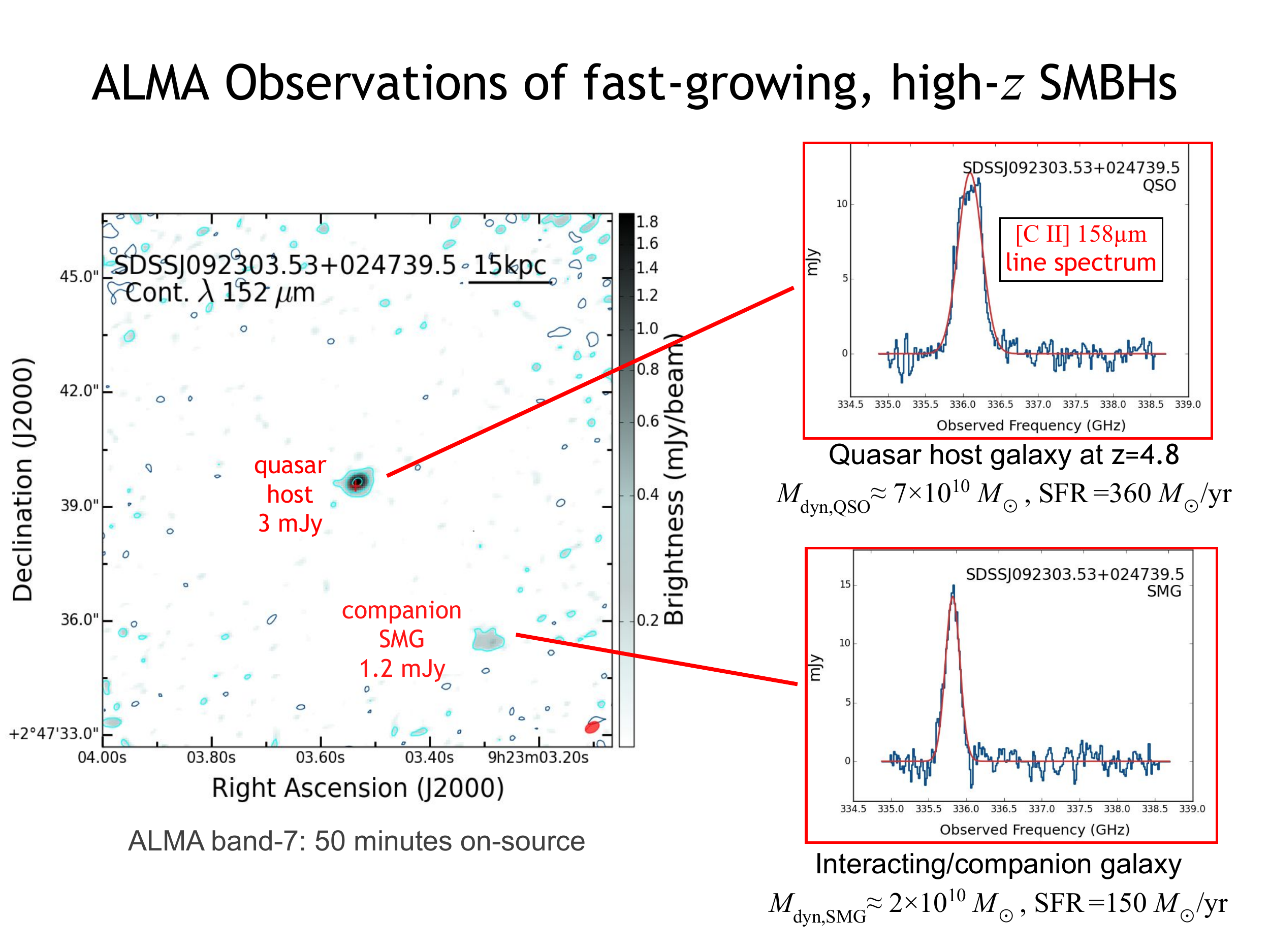}
    \caption{ALMA allows to study the host galaxies and larger-scale environments of $z\gtrsim5$ quasars (taken from \cite[Trakhtenbrot \et 2017b]{Trakhtenbrot2017_z48_ALMA}). 
    {\it Centre:} A 50-minute integration with ALMA band-7 allows to robustly detect the cold ISM, and determine the SFR, in the host galaxy of a $z\simeq5$ quasar. 
    {\it Left:} The spatially-resolved \CII\ emission suggests rotation-dominated dynamics, and allows to estimate the host dynamical mass.
    {\it Right:} The same data cube reveals a dusty, star-forming  companion galaxy, separated from the quasar by $\sim$40 kpc and $\sim$250 \kms.
    }
    \label{fig:ALMA_host_and_env}
\end{figure*}

\smallskip
Importantly, sub-mm observations also probe several ISM emission lines, most notably [C\,{\sc ii}]\ $\lambda$157.74 $\mu{\rm m}$, which allow to study the dynamics of the quasar hosts. 
Since the first detection of \cii\ in the $z\simeq6.4$ quasar J1148+5251 (\cite[Maiolino \et 2005]{Maiolino2005_CII_J1148}), such measurements have grown to become a booming ``industry'', with several teams pursuing increasingly better data for ever growing samples 
(for an impression of this progress, see, e.g.,
\cite[Venemans \et 2012]{Venemans2012_z71_CII}, 
\cite[2016]{Venemans2016_z6_cii}, 
\cite[2017]{Venemans2017_z75_CII}, 
\cite[2019]{Venemans2019_z66_ALMA_hires}; 
\cite[Wang \et 2013]{Wang2013_z6_ALMA}, 
\cite[2019b]{Wang2019_z6_ALMA_hires}
\cite[Willott \et 2013]{Willott2013_z6_ALMA}, 
\cite[2015]{Willott2015_CFHQS_ALMA}, 
\cite[2017]{Willott2017_z6_SFR}; 
\cite[Trakhtenbrot \et 2017b]{Trakhtenbrot2017_z48_ALMA}, 
\cite[Decarli \et 2018]{Decarli2018_CII_z6}, 
\cite[Izumi \et 2018]{Izumi2018_SHELLQs_III_z6_ALMA}, 
\cite[2019]{Izumi2019_SHELLQs_VIII_z6_ALMA}).
In many cases, the spatially resolved \cii\ line velocity maps show signs of ordered, rotation-dominated gas dynamics (Fig.~\ref{fig:ALMA_host_and_env}, left; see, e.g., \cite[Trakhtenbrot \et 2017b]{Trakhtenbrot2017_z48_ALMA}, \cite[Shao \et 2017]{Shao2017_ALMA_z6}).
This, in turn, provides further motivation to use the velocities and spatial extent of the line emitting regions as probes of the {\it dynamical} host masses.
It's important to note, however, that this common practice necessitates several crucial assumptions, mainly that of a rotating cold gas disk, and thus carries significant uncertainties.
Indeed, deeper and higher resolution ALMA data obtained for some systems show no (or very limited) evidence for rotation-dominated gas structures, thus highlighting the limitations of such assumptions. 
Examples of such cases, and of more elaborate analyses, can be found in, e.g., \cite[Neeleman \et (2019)]{Neeleman2019_z6_comp_ALMA}, \cite[Venemans \et (2019)]{Venemans2019_z66_ALMA_hires}, and \cite[Wang \et (2019b)]{Wang2019_z6_ALMA_hires}.

\smallskip
The rough dynamical host mass estimates derived from \cii\ line measurements are typically of order $\mdyn\sim10^{10-11}\,\Msol$ (e.g., \cite[Willott \et 2015]{Willott2015_CFHQS_ALMA}, \cite[Venemans \et 2016]{Venemans2016_z6_cii}, \cite[Shao \et 2017]{Shao2017_ALMA_z6}, \cite[Trakhtenbrot \et 2017b]{Trakhtenbrot2017_z48_ALMA}, \cite[Decarli \et 2018]{Decarli2018_CII_z6}, \cite[Izumi \et 2019]{Izumi2019_SHELLQs_VIII_z6_ALMA}). 
The {\it stellar} masses are naturally somewhat lower, although they would most likely remain inaccessible at least until \jwst\ is operating.

\begin{figure*}
    \centering
    \includegraphics[trim={0 0cm 2cm 2.5cm},clip,width=0.475\textwidth]
    {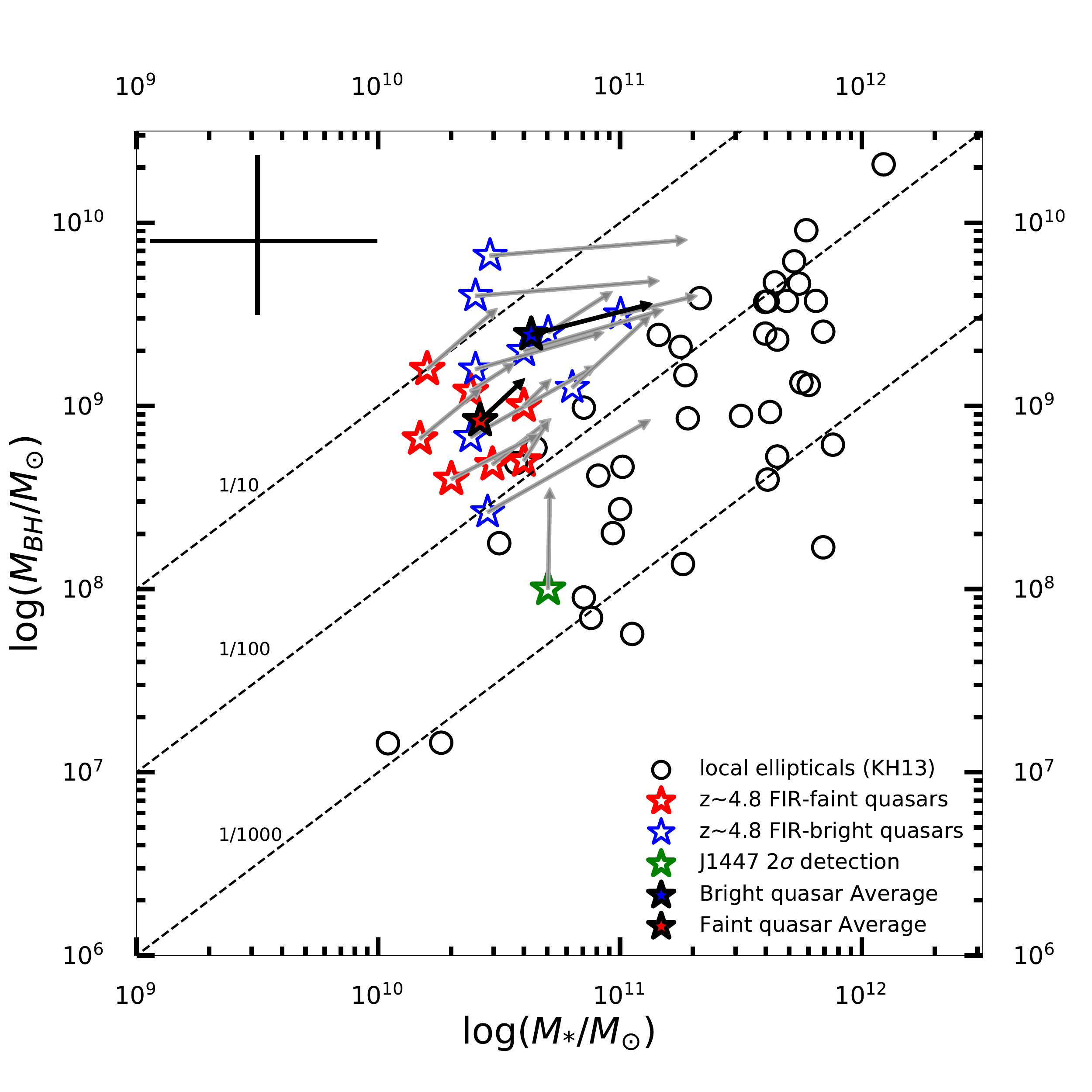}
    \hspace{0.8cm}
    \includegraphics[trim={0 -0.55cm 0cm 0cm},clip,width=0.425\textwidth]
    {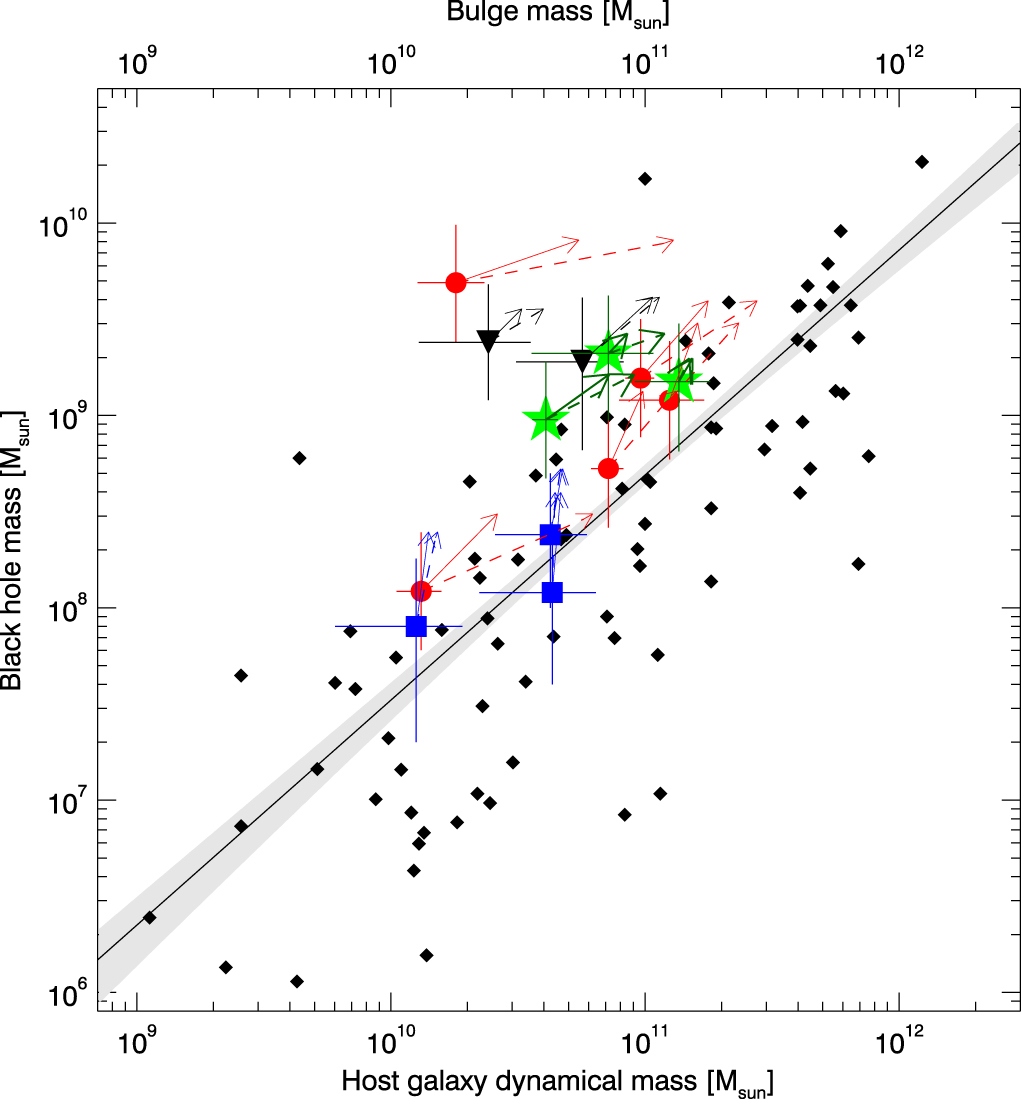}
    \caption{The SMBH-vs-host mass plane for $z\gtrsim5$ quasars, with ALMA-enabled measurements of host {\it dynamical} (gas+stellar) masses, \mdyn.
    The two panels show measurements for quasars at $z\simeq5$ (left; taken from \cite[Nguyen \et 2020]{Nguyen2020_z48_ALMA}) and at $z\sim6$ (right; measurements compiled by \cite[Venemans \et 2016]{Venemans2016_z6_cii}. 
    In both panels, the coloured symbols mark the high-$z$ quasars, which have large systematic uncertainties (large cross in left panel), and the black symbols mark local, inactive systems (taken from \cite[Kormendy \& Ho 2013]{KormendyHo2013_MM_Rev}).
    For each high-$z$ quasar, an arrow illustrates the expected evolution of the system within 50 Myr, assuming the measured \Lagn\ (${\propto}\Mbhdot$) and SFR (${\approx}\dot{M}_{\rm host}$).
    The highest redshift quasars appear to have higher $\mbh/\mhost$ than what is seen in the local universe, with $\mbh/\mhost\sim1/100-1/30$. 
    Their short-term subsequent evolution may bring them somewhat closer to the locally observed SMBH-host relation.
    }
    \label{fig:MM}
\end{figure*}

\smallskip
Compared with the SMBH-host relations seen in the local Universe, and based on the highly uncertain \mdyn\ estimates, the highest-redshift quasars tend to have somewhat over-massive SMBHs, with SMBH-to-host mass ratios of $\mbh/\mhost\sim1/100-1/30$ (Fig.~\ref{fig:MM}). 
This provides some evidence in support of the idea that the first generation of SMBHs emerged through some sort of preferentially efficient BH fuelling mechanism. 
However, as with lower-redshift systems, it's important to keep in mind that the most luminous $z\gtrsim5$ quasars currently studied may be biased towards high $\mbh/\mhost$, while lower-mass systems could be closer to the local $\mbh/\mhost$ ratio (or even below it). 
Indeed, new and deep ALMA data obtained for (relatively) lower-luminosity $z\sim6$ quasars from the SHELLQs sample revealed $\mbh/\mhost$ ratios consistent with the local ones (\cite[Izumi \et 2019]{Izumi2019_SHELLQs_VIII_z6_ALMA}).
Some studies have tried to go further, and estimated the (short-term) evolution of \mbh\ and \mhost\ in some $z\gtrsim5$ quasars, based on the measured \Lagn\ and SFRs. 
Most systems, and particularly the highest $\mbh/\mhost$ ones, seem to be able to get closer to the locally-observed mass relation (see arrows in Fig.~\ref{fig:MM}).

To summarise, the best data available for the highest-redshift quasars currently known do {\it not} show overwhelming evidence for large, systematic and robust deviations from the locally-observed $\mbh-\mhost$ relations.

\subsection{Outflows}
\label{subsec:outflows}

The radiative and/or mechanical energy output of accreting SMBHs onto their host galaxies, so-called ``AGN feedback'', is a widely-acknowledged cornerstone of the co-evolutionary paradigm. 
Specifically, given the requirement for (nearly) continuous BH growth at high rates, one could expect the highest-redshift quasars to showcase such feedback processes. 
Combined with the short time available for host evolution, these systems may be considered as optimal test-beds for any viable feedback scenario.

\begin{figure*}
    \centering
    \includegraphics[width=0.85\textwidth]
    {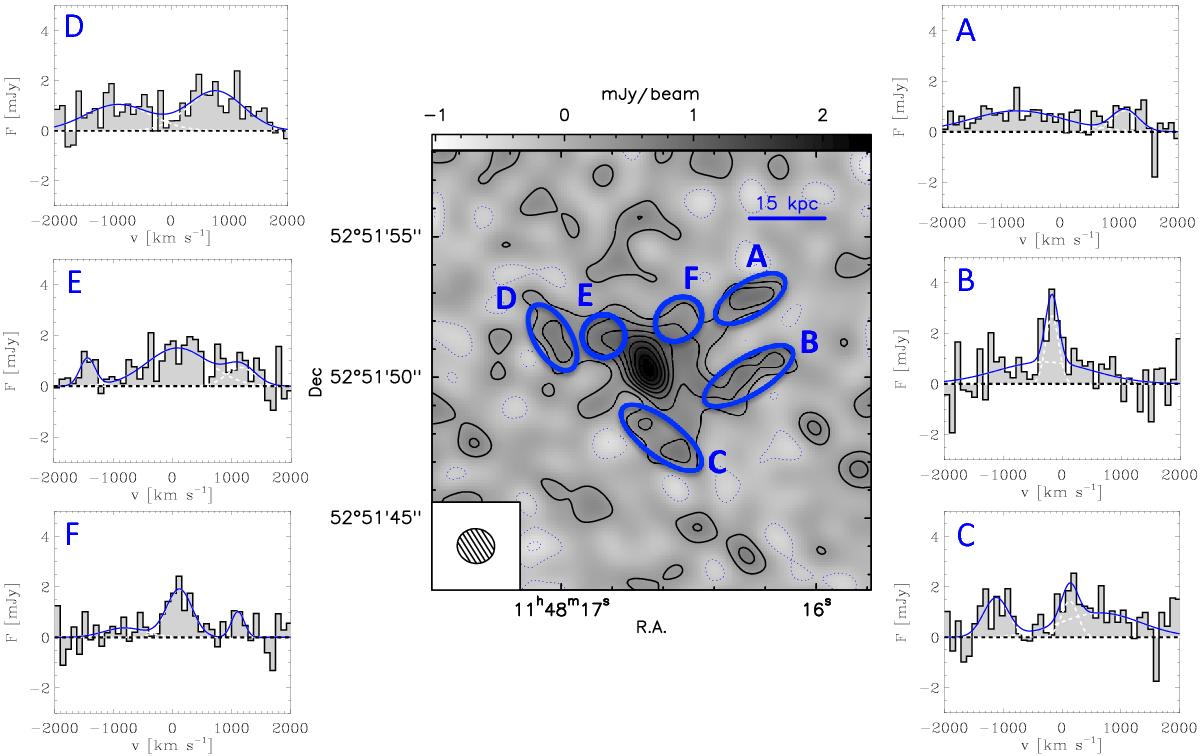}
    \caption{A large-scale outflow of cold seen in a $z\simeq6.4$ quasar (taken from \cite[Cicone \et 2015]{Cicone2015_J1148}). 
    The outflow, identified through a Plateau de Bure Interferometer observation of broad \CII\ emission components, extends out to $>$20 kpc and $>$1000 \kms. 
    }
    \label{fig:outflow_cicone}
\end{figure*}

\smallskip
Currently, the only evidence for anything that may be interpreted as AGN feedback comes in the form of extended, high-velocity outflows of cold gas, seen in some of the highest-redshift quasars.
In particular, \cite{Maiolino2012_J1148_feedback}, and later \cite{Cicone2015_J1148}, have identified \& resolved such an outflow, extending out to $>$15 kpc and reaching $>$1,000 \kms, using PdBI observations of the \cii\ line in a $z=6.4$ quasar (Fig.~\ref{fig:outflow_cicone}).

More recently, two different studies used {\it stacking} analysis of \cii\ data for dozens of $z\gtrsim5$ quasars to try and identify such outflow signatures, through broad (but weak) \cii\ emission components.
\cite{Bischetti2019_ALMA_stack} studied 48 quasars at $4.5 < z < 7.1$ and found broad \cii\ wings that are interpreted to trace  outflowing gas with velocities $v\gtrsim$1,000\,\kms, an extent of $\gtrsim$5 kpc, and a deduced a mass outflow rate of order $\sim100\,\mpyr$. 
On the other hand, \cite{Stanley2019_ALMA_z6_stack} found only marginal evidence (${<}3\sigma$) for such outflow signatures in their analysis of 26 $z\sim6$ quasars (the outflow signatures may be more prominent in a subset of about half of these quasars).

\smallskip
Given the high SFRs and rich gas content of the highest-redshift quasar hosts, it may not seem straightforward to link the large-scale, energetic outflows probed in broad \cii\ emission directly to AGN activity.
However, there is strong evidence, from low-redshift systems, that the extremely high velocity regime ($\gtrsim$1,000 \kms) tends to be indeed directly linked to AGN-driven outflows (see, e.g., \cite[Veilleux \et 2013]{Veilleux2013_Herschel_outflows}, \cite[Janssen \et 2016]{Janssen2016_CII_outflows}, \cite[Stone \et 2016]{Stone2016_PACS_AGN_outflows}).
Future studies of high-ionization (rest-frame) optical emission lines (e.g., \OIII, \NIIopt), using \jwst/NIRSpec 3D spectroscopy on $\sim$1 kpc scales, would allow us to search for more direct signatures of AGN-driven outflows.

\subsection{Large-scale environments}
\label{subsec:env}

Several models suggest that the first generation of SMBHs, and indeed the most luminous quasars at $z\gtrsim5$, would be found in over-dense large-scale environments (see, e.g., \cite[Overzier \et 2009]{Overzier2009}, \cite[Costa \et 2014]{Costa2014_z6_env_sims}, \cite[Habouzit \et 2019]{Habouzit2019_z6_env}).
The general idea is that in such regions, which could later evolve into rich galaxy (proto-)clusters, IGM gas can be efficiently funnelled into the central galaxies, and thus to the SMBHs at their hearts.

\smallskip
Attempts to address these ideas observationally, using multi-band optical imaging designed to identify LBGs at redshifts comparable to those of the quasars themselves, have so far resulted in rather ambiguous results. 
While some studies claimed to identify over-dense environments (e.g., \cite[Overzier \et 2006]{Overzier2006_z52}, \cite[Kim \et 2009]{Kim2009_idrops_z6}, \cite[Utsumi \et 2010]{Utsumi2010_env_J2329}, \cite[Husband \et 2013]{Husband2013}), other systems show no evidence for such over-densities (e.g., \cite[Willott \et 2005]{Willott2005_z6_comp}, \cite[Ba{\~n}ados \et 2013]{Banados2013_z57_env}, \cite[Simpson \et 2014]{Simpson2014_ULASJ1120_env}). 
Such observational efforts are limited by their focus on (rest-frame) UV bright LBGs; by the coarse redshift resolution they provide for any nearby source ($\Delta z{\sim}0.5$); by cosmic variance; and by the relatively long exposure times they require. 
Linking observations to models is further complicated by the wide range in BH seeding scenarios; by the possible effects of AGN feedback; and other details (see discussion in \cite[Buchner \et 2019]{Buchner2019_z6_syn_model} and \cite[Habouzit \et 2019]{Habouzit2019_z6_env}).

\smallskip
As with host galaxy studies, here too ALMA is revolutionising our understanding of the highest-redshift quasars.
The same ALMA observations that probe the quasar hosts can also be used to search for real ``companion'' galaxies (i.e., not just projected neighbouring sources), separated from the quasars by up to $\sim$50 kpc and/or $\sim$500\,\kms. 
The occurrence rate of such companion galaxies is surprisingly high.
\cite{Trakhtenbrot2017_z48_ALMA} found three such companions among a sample of six $z\simeq4.8$ quasars, and a follow-up study of 12 additional systems from the same parent sample brings the total to 5 companions among 18 quasars (i.e., 28\%; \cite[Nguyen \et 2020]{Nguyen2020_z48_ALMA}).
\cite{Decarli2017_z6_ALMA_comp} identified 4 companion galaxies among their sample of 25 quasars at $z\sim6$ (i.e., 16\%).
\cite{Izumi2019_SHELLQs_VIII_z6_ALMA} identified one companion galaxy, and another candidate companion, among a sample of merely three $z\sim6$ quasars drawn from the SHELLQs project.
Fig.~\ref{fig:ALMA_host_and_env} illustrates a quasar+companion system, from the \cite{Trakhtenbrot2017_z48_ALMA} study.
These occurrence rates are far higher than what is seen among normal, SF galaxies at $z\sim5-7$ observed in deep extragalactic surveys (i.e., LBGs; e.g., \cite[Aravena \et 2016]{Aravena2016_HUDF_cii}).

\smallskip
The companion galaxies detected so far share a few interesting properties.
First, their detection with ALMA immediately implies that they are themselves gas-rich and have intense SF activity, with SFRs of order 100 \mpyr.
Second, their continuum emission and (rough) dynamical masses are within a factor of $\sim$3 from those of the quasar hosts. 
This, along with other observed similarities (e.g., \cite[Walter \et 2018]{Walter2018_OIII_88mic_z6}), suggests that these systems may be considered as major galaxy mergers between rather similar galaxies.
Finally, and crucially, the companion galaxies are not detected in (deep) NIR imaging, including with \hst\ (\cite[Decarli \et 2017]{Decarli2017_z6_ALMA_comp}, \cite[Mazzucchelli \et 2019]{Mazzucchelli2019_z6_comp}), suggesting they are indeed extremely dusty, perhaps similar to well-known merging galaxies in the local Universe (e.g., Arp~220; see \cite[Mazzucchelli \et 2019]{Mazzucchelli2019_z6_comp}).
This latter point may also explain why the aforementioned searches for neighbouring LBGs have not identified a similarly high occurrence rate of close companions and/or signs of over-dense environments. 

\smallskip
Linking the ALMA-detected companions to any detailed scenario for the emergence of the earliest quasars remains challenging.
First, our understanding of \cii-based selection of (inactive) $z\gtrsim5$ galaxies -- that is, the nature and abundance of galaxies comparable to the quasar companions -- is itself limited, and fast-changing (see, e.g., \cite[Capak \et 2015]{Capak2015_CII_COSMOS}, \cite[Aravena \et 2016]{Aravena2016_HUDF_cii}, \cite[Faisst \et 2019]{Faisst2019_ALPINE}).
Second, we cannot know how long would any of the observed galaxy-galaxy interactions last.
Moreover, while the galaxy interactions we see may perhaps explain the triggering of the concurrent quasar phase, they could {\it not} have powered the entire period of nearly continuous SMBH growth, required to account for the high \mbh\ we measure. 
Given the typical timescales of galaxy-galaxy interactions, it is not unreasonable to think that the highest-redshift quasars experienced a sequence of major mergers, and we are able to witness the early, rather prolonged phases of some of these mergers, in some of the systems.
In this context, perhaps the highest-SFR quasar hosts, which are generally {\it not} those with $>$10 kpc companions, may be powered by advanced (late-stage), unresolved major mergers (see also \cite[Izumi \et 2019]{Izumi2019_SHELLQs_VIII_z6_ALMA} for a candidate advanced merger system).
\cite[Trakhtenbrot \et (2017b)]{Trakhtenbrot2017_z48_ALMA} offers a rather detailed discussion of both the prospects, and limitations, of linking the (ALMA-detected) quasar companions to the scenario of merger-driven growth for the first generation of SMBHs.

\smallskip
Given this significant progress, as well as other exciting results at somewhat lower redshifts (e.g., \cite[Banerji \et 2017]{Banerji2017_redQSOs_ALMA}, \cite[Bischetti \et 2018]{Bischetti2018_z44_merger}, \cite[Miller \et 2018]{Miller2018_SPT_cluster_z43}), ALMA will likely continue to be the main surveyor of the environments of high-redshift quasars (together with other large sub-mm facilities).

\section{Where are the Lower-Luminosity, Lower-Mass Counterparts?}
\label{sec:lowL}

We are clearly gaining various new insights into the nature of luminous, unobscured quasars at $z\gtrsim5$, powered by highly accreting SMBHs with $\mbh\gtrsim10^{8}\,\Msol$. 
However, one must recall that such systems are extremely rare, with space densities are of order of $\Phi\simeq10^{-7}\,{\rm Mpc}^{-3}$ (for $\Lagn\gtrsim10^{46}\,\ergs$; e.g., \cite[Kulkarni \et 2019]{Kulkarni2019_QLF}, \cite[Wang \et 2019]{Wang2019_z65_QLF}, \cite[Shen \et 2020]{Shen2020_QLF}). 
As such, they are expected to represent only the tip of the iceberg of the entire population of (active) SMBHs at these early cosmic epochs. 
What do we know, observationally, about the lower-\Lagn, lower-\mbh\ AGN at $z\gtrsim5$? 
The answer is, perhaps surprisingly, ``very little''.

\begin{figure*}
    \centering
    \includegraphics[height=0.275\textheight]
    {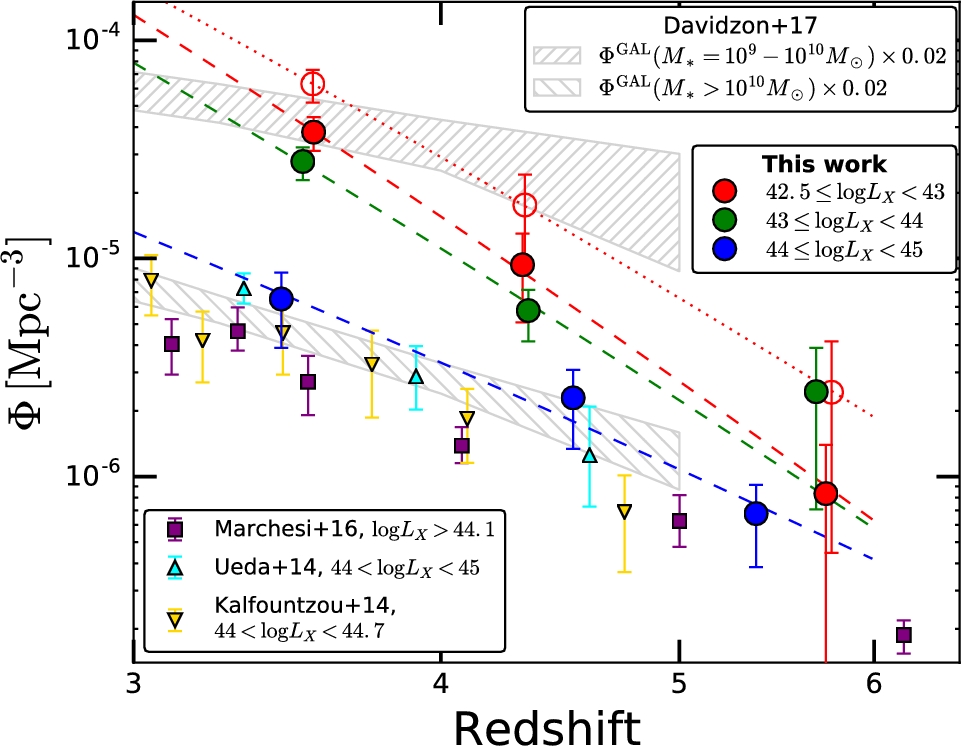}
    \hfill 
    \includegraphics[trim={0 -1.0cm 0 0},clip,height=0.25\textheight]
    {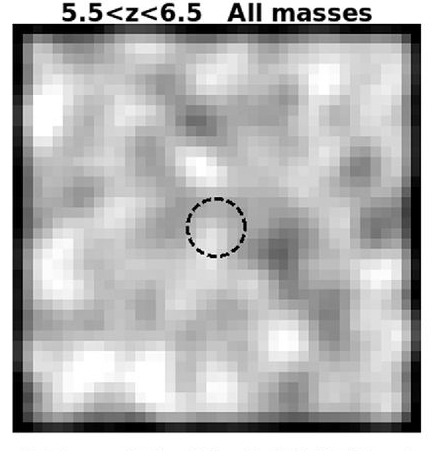}
    \caption{The ``missing'' population of $z\gtrsim5$ low-luminosity, low-\mbh\ AGN.
    {\it Left:} the decreasing space density of X-ray selected, medium- and low-$L$ AGN, towards high redshifts, as seen in deep multi-wavelength surveys (mainly CDF-S, \cite[Luo \et 2017]{Luo2017_CDFS_7Ms}, and COSMOS, \cite[Marchesi \et 2016]{Marchesi2016_XVP_hiz}).
    Note the lack of robustly identified, spectroscopically confirmed, $z\gtrsim5-6$ AGN (figure taken from  \cite[Vito \et 2018]{Vito2018_hiz_XLF_fobs}).
    {\it Right:} the lack of AGN-related X-ray emission in the deepest available {\it stack} of X-ray data for abundant, (rest-frame) UV-bright SF galaxies at $z\sim6$ (LBGs; figure taken from \cite[Vito \et 2016]{Vito2016_CDFS_7Ms}).
    This X-ray stack reaches a depth of $>$1 {\it Giga}-second, and shows that the typical AGN luminosity in such galaxies is $L_{\rm AGN} \lesssim 5{\times}10^{43}\,\ergs$.
    }
    \label{fig:lowL}
\end{figure*}

\smallskip
First, there is clear evidence for a drastic decline in the space densities of AGN at $z\sim3-5$, and particularly for the lower-luminosity ones, as traced by the deepest, highly complete X-ray and multi-wavelength surveys (Fig.~\ref{fig:lowL}, left; see, e.g., \cite[Marchesi \et 2016]{Marchesi2016_XVP_hiz}, \cite[Luo \et 2017]{Luo2017_CDFS_7Ms}, and \cite[Vito \et 2018]{Vito2018_hiz_XLF_fobs} for the most up-to-date deep surveys). 

\smallskip
Second, these same surveys have not (yet) provided robust identifications of $z\gtrsim5.5$ AGN, despite intensive, dedicated efforts (see, e.g., \cite[Weigel \et 2015]{Weigel2015_CDFS} and \cite[Cappelluti \et 2016]{Cappelluti2016_CANDELS_S}, and the compilation in \cite[Vito \et 2018]{Vito2018_hiz_XLF_fobs}; these contrast the findings of \cite[Giallongo \et 2015]{Giallongo2015_CANDELS_hiz_AGN}).
Specifically, there is currently only one spectroscopically confirmed $z{>}5$ AGN in the {\it Chandra} Deep Fields (CDFs; that source is at $z{=}5.19$, see \cite[Vito \et 2018]{Vito2018_hiz_XLF_fobs}), and no AGN beyond $z\simeq5.2$ in neither the two CDFs nor in the COSMOS field (\cite[Marchesi \et 2016]{Marchesi2016_XVP_hiz}).
Attempts to corroborate additional candidates (i.e., with high {\it photometric} redshifts) are still on-going, using exceedingly deep spectroscopic obseravations on large telescopes.
It should be noted that the dearth of detections at $z\gtrsim5$ is {\it not} due to insufficient survey volume or insufficient sensitivity (see Fig.~6 in \cite[Trakhtenbrot \et 2016]{Trakhtenbrot2016_COSMOSFIRE_MBH} and the discussion in \cite[Weigel \et 2015]{Weigel2015_CDFS}).

\smallskip
Finally, {\it stacking} analysis of the deepest X-ray data available for hundreds of $z\gtrsim5$ galaxies, using the CDF-S 7 Ms data -- and reaching the remarkable effective depth of $\sim$1 {\it Giga}-second -- resulted in no robust detection of X-ray emission (Fig.~\ref{fig:lowL}, right; \cite[Vito \et 2016]{Vito2016_CDFS_7Ms}).
This implies that the typical AGN-related emission in abundant $z\gtrsim5$ galaxies has $\Lagn \lesssim 5{\times}10^{43}\,\ergs$.

\smallskip
The reason behind this apparent dearth of lower-luminosity AGN signatures at $z\gtrsim5$ is unclear. 
It could be driven by either 
(1) an exceptionally high fraction of highly obscured systems; 
(2) exceptionally low radiative efficiencies (i.e., ADAF/RIAF-like accretion); 
(3) extremely low duty cycles (i.e., due to ``flickering''); 
(4) a low BH occupation fraction in such galaxies; 
and/or a combination of these scenarios.
Given this wide range of possibilities, any attempt to compare evolutionary models with observations (including by future facilities; see \S\ref{sec:future} below) is complicated by degeneracies in the relevant model parameters.
%

\smallskip
It should finally be noted that the apparent lack of lower-luminosity AGN emission at $z\gtrsim5$ is not (yet) in too great of a tension with neither {\it all} models, nor with {\it all} sensible extrapolations of the quasar luminosity function (compare, e.g., Fig.~16 in \cite[Vito \et 2016]{Vito2016_CDFS_7Ms} with Fig.~6 in \cite[Kulkarni \et 2019]{Kulkarni2019_QLF} or Fig.~5 in \cite[Shen \et 2020]{Shen2020_QLF}).

\smallskip
Clearly, yet deeper (X-ray) surveys are needed to detect low-luminosity, low-mass $z\gtrsim5$ SMBHs, or otherwise to constrain their abundance, and to directly confront the wide-ranging models for their emergence. 
The data in hand already motivates us to consider some intriguing scenarios.

\section{Future Prospects}
\label{sec:future}

The near future holds several opportunities to greatly improve our understanding of the currently-known $z\gtrsim5$ quasar population and, importantly, to reveal new and much larger populations of early SMBHs.

\smallskip
Once operating, \jwst\ will be able to directly probe the stellar and ionized gas components in the hosts of the already-known quasars. 
This can be done through NIR and MIR imaging, and NIR integral field spectroscopy.
Several theoretical studies have already provided predictions for what \jwst\ will be able to see, particularly in terms of the stellar emission (e.g., \cite[Natarajan \et 2017]{Natarajan2017_JWST_model}, \cite[Volonteri \et 2017]{Volonteri2017_JWST_model}).
This will hopefully allow us to better understand early SMBH-host co-evolution, and perhaps even constrain high-mass BH seeding scenarios. 
Another improvement will come from the ability to observe the broad Balmer emission lines, which provide the most reliable kind of ``virial'' \mbh\ estimates. 
It is important to keep in mind that \jwst\ is not designed to be a survey/discovery facility (at least in the context of high-$z$ quasars), thus careful consideration of the kind of targets and samples to be observed, and the possible biases they entail, is key.

\smallskip
There are several upcoming and future facilities that will survey and discover {\it new} accreting SMBHs at $z\gtrsim5$.
In the optical/NIR regime, {\it Euclid} is predicted to be able to detect dozens of highly (UV) luminous quasars, with $\Lbol\gtrsim10^{47}\,\ergs$ -- similar to SDSS $z\sim6$ quasars, but out to $z\sim8-10$ (i.e., $\sim$0.5 Gyr after the Big Bang).
The wide-field surveys planned to be conducted with the LSST and with {\it WFIRST} will be able to uncover somewhat  lower-luminosity quasars at comparably early epochs, reaching $\Lbol\simeq10^{46}\,\ergs$ (i.e., driven by Eddington-limited SMBHs with $\mbh\simeq10^8\,\Msol$, or more massive, slowly accreting systems).
Future X-ray missions hold the key to revealing {\it all} accreting SMBHs at these epochs, regardless of obscuration.
{\it Athena} is planned to discover {\it hundreds} of X-ray emitting AGN beyond $z\sim6$ with yet lower luminosities, reaching $\Lbol\simeq10^{45}\,\ergs$.
Ultimately, one has to pursue much deeper, high resolution X-ray observations to directly probe the fast growth of newly formed SMBHs.
The future {\it Lynx} X-ray mission concept is envisioned to have about $\times$100 the sensitivity of {\it Chandra}, allowing to discover AGN as faint as  $\Lbol\simeq10^{42}\,\ergs$ out to $z\sim15$, in a 4 Ms ``deep drill'' survey.
This has the potential of detecting SMBHs with masses as low as $\mbh\simeq10^{4}\,\Msol$ at the epoch of massive BH seeding (see magenta line in Fig.~\ref{fig:MBH_vs_z}), or otherwise put strong constraints on their existence. 

\smallskip
Future facilities and surveys will thus be able to directly probe the emergence of the first SMBHs in the Universe, within the first hundreds of Myr after the Big Bang. 
It is important, however, to keep in mind that we will also need extensive, multi-wavelength follow-up capabilities to maximize the science return from these newly-discovered SMBH populations, and to more critically test the relevant models.


\vspace*{-0.5cm}
\section{Summary}
\label{sec:summary}

Table~\ref{tab:props} summarises the key observed and derived properties of the known highest redshift quasars, at $z\sim5-7$, as well as what would be (naively) expected for their lower-luminosity, lower-\mbh\ counterparts, which are not yet robustly identified. 

\begin{table}
  \begin{center}
  \caption{Overview of current observed and derived properties of the highest-redshift quasars, and expectations for their yet-to-be established lower-mass counterparts.}
  \label{tab:props}
 {\scriptsize
  \begin{tabular}{|l|c|c|}\hline 
{\bf Property} & {\bf Known $z\sim5-7$} & {\bf ``Typical'' AGN /} \\ 
~~~            & {\bf quasars}          & {\bf galaxies} \\ 
\hline 
Luminosity, \Lbol       & ${\gtrsim}10^{46}\,\ergs$   & ${\lesssim}10^{45}\,\ergs$  \\
Obscuration / selection & un-obscured / UV-opt.    & ${\sim}$50\% obscured / X-ray \\ 
\hline 
SMBH mass, \mbh         & ${\sim}10^{9}\,\Msol$      & ${\sim}10^{7}\,\Msol$ \\    
Accretion rate, \lledd  & ${\sim}$1                  & ${\sim}0.01-1$     \\    
Accretion mode          & thin disk, $\eta{\gtrsim}0.1$ & (who knows, really?) \\
\hline 
Implied BH seeds        & massive,                      & stellar (pop-III),   \\
~~~                     & $\mseed{\sim}10^{4-6}\,\Msol$ & $\mseed{<}10^{3}\,\Msol$ \\
\hline 
Host mass, \mhost       & ${\sim}10^{10-11}\,\Msol$  & ${\sim}10^{9-10}\,\Msol$ \\
Host SFR                & ${\sim}100-3000\,\mpyr$    & ${<}100\,\mpyr$       \\
\hline 
Large-scale env.        & over-dense, mergers, outflows & ``normal''?          \\
\hline 
Demographics            & rare! $\Phi{\lesssim}10^{-7}\,{\rm Mpc}^{-3}$ & common? $\Phi{\gtrsim}10^{-5}\,{\rm Mpc}^{-3}$ \\
~~~~ & ~~~~ & ($\sim$10\% of galaxies? less?) \\
\hline 
Future prospects        & {\it Euclid}, {\it Athena}, {\it WFIRST} & {\it Lynx} \\
\hline 
\end{tabular}
}
\end{center}
\end{table}

\vspace*{0.25cm} 
\noindent
The key take-away points from this short overview can be summarized as follows: 
\vspace*{0.25cm} 

\begin{itemize}
    
    \item By now, there are hundreds of highly luminous, unobscured AGN (quasars) known at $z\gtrsim5$. 
    Their basic observed properties in the AGN-dominated, X-ray-to-NIR regime do not differ from their lower-$z$ counterparts. 
    We may, however, expect these early systems to stand out in {\it some} way, which can be linked to the fast growth of the SMBHs that power them, within the first Gyr after the Big Bang.

    \item 
    ALMA is revolutionising our ability to study the host galaxies, and larger-scale environments, of the highest-redshift quasars. 
    The hosts are massive, gas-rich galaxies with a wide range of SFRs, reaching the highest levels known ($\sim3000\,\mpyr$).
    
    \item 
    ALMA allowed us to uncover dusty, star-forming galaxies accompanying, and indeed interacting with, the quasar hosts. 
    The occurrence rate of such companions is much higher than what is seen in inactive high-$z$ galaxies, suggesting that early, fast SMBH growth may be linked to over-dense cosmic environments and/or galaxy mergers.

    \item 
    We have yet to robustly identify the lower-luminosity, lower-\mbh\ AGN population at $z\gtrsim5$, despite the remarkably deep X-ray survey data currently in hand. 
    Such systems are expected to be associated with (a fraction of) the much more common, ``normal'' SF galaxies at these early epochs. 
    This dearth of AGN signatures may be driven by a high fraction of highly obscured systems; by low radiative efficiencies; by a low duty cycle; and/or a low BH occupation fraction. 

    \item 
    Upcoming and future facilities and surveys will allow us to directly probe the stars and gas in the hosts of the currently known $z\gtrsim5$ quasars (with \jwst); 
    to detect many more highly luminous quasars out to $z\sim10$ (with LSST \& {\it WFIRST}), 
    as well as hundreds of obscured, lower-luminosity systems (with {\it Athena}). 
    The {\it Lynx} future mission concept paves the way to directly probe the epoch of massive BH seed formation.

\end{itemize} 

\vspace*{0.5cm}
\noindent
With growing samples and ever-improving data, our understanding of the highest-redshift quasars is now better than ever. 
However, to fully address the sophisticated models that were developed to explain the first generation of massive BHs in the universe, we have to strive for deeper, more detailed observations with new facilities and surveys.

\bigskip
\noindent
\textit{
I thank the organisers of the meeting for their kind invitation to present this overview, for the opportunity to take part in this exciting meeting, and indeed to visit Ethiopia and get to familiar with the local scientific community.
I also thank the meeting participants, particularly N.~Brandt and H.~Netzer, for their insightful comments following my presentation, which helped me to improve this written contribution.
} 
 







\begin{thebibliography}{}

\bibitem[Agarwal et al. (2013)]{Agarwal2013_obese} 
Agarwal, B., Davis, A.~J., Khochfar, S., et al.\ 2013, \textit{MNRAS}, 432, 3438

\bibitem[Aravena et al. (2016)]{Aravena2016_HUDF_cii} 
Aravena, M., Decarli, R., Walter, F., et al.\ 2016, \textit{ApJ}, 833, 71

\bibitem[Banerji et al. (2017)]{Banerji2017_redQSOs_ALMA} 
Banerji, M., Carilli, C.~L., Jones, G., et al.\ 2017, \textit{MNRAS}, 465, 4390

\bibitem[Belladitta et al.(2019)]{Belladitta2019_z5_blazar_DES} 
Belladitta, S., Moretti, A., Caccianiga, A., et al.\ 2019, \textit{A\&A}, 629, A68

\bibitem[Ba\~{n}ados et al. (2013)]{Banados2013_z57_env} 
Ba{\~n}ados, E., Venemans, B., Walter, F., et al.\ 2013, \textit{ApJ}, 773, 178

\bibitem[Ba{\~n}ados et al. (2016)]{Banados2016_PS1_z6_qsos} 
Ba{\~n}ados, E., Venemans, B.~P., Decarli, R., et al.\ 2016, \textit{ApJS}, 227, 11

\bibitem[Ba{\~n}ados et al. (2018a)]{Banados2018_z75_Nature} 
Ba{\~n}ados, E., Venemans, B.~P., Mazzucchelli, C., et al.\ 2018a, \textit{Nature}, 553, 473

\bibitem[Ba{\~n}ados et al. (2018b)]{Banados2018_z75_Xray} 
Ba{\~n}ados, E., Connor, T., Stern, D., et al.\ 2018b, \textit{ApJL}, 856, L25

\bibitem[Ba{\~n}ados et al. (2018c)]{Banados2018_radio_z6} 
Ba{\~n}ados, E., Carilli, C., Walter, F., et al.\ 2018c, \textit{ApJL}, 861, L14

\bibitem[Bischetti et al. (2018)]{Bischetti2018_z44_merger} 
Bischetti, M., Piconcelli, E., Feruglio, C., et al.\ 2018, \textit{A\&A}, 617, A82

\bibitem[Bischetti et al. (2019)]{Bischetti2019_ALMA_stack} 
Bischetti, M., Maiolino, R., Carniani, S., et al.\ 2019, \textit{A\&A}, 630, A59

\bibitem[Buchner et al. (2019)]{Buchner2019_z6_syn_model} 
Buchner, J., Treister, E., Bauer, F.~E., et al.\ 2019, \textit{ApJ}, 874, 117

\bibitem[Capak et al. (2015)]{Capak2015_CII_COSMOS} 
Capak, P.~L., Carilli, C., Jones, G., et al.\ 2015, \textit{Nature}, 522, 455

\bibitem[Cappelluti et al. (2016)]{Cappelluti2016_CANDELS_S} 
Cappelluti, N., Comastri, A., Fontana, A., et al.\ 2016, \textit{ApJ}, 823, 95

\bibitem[Carnall et al. (2015)]{Carnall2015_z6_qsos} 
Carnall, A.~C., Shanks, T., Chehade, B., et al.\ 2015, \textit{MNRAS}, 451, L16

\bibitem[Casey {et~al.} (2014)]{Casey2014_SMGs_rev}
Casey, C.~M., Narayanan, D., \& Cooray, A.~R. 2014, \textit{Phys. Rep.}, 541, 45

\bibitem[Cicone {et~al.} (2015)]{Cicone2015_J1148}
Cicone, C., Maiolino, R., Gallerani, S., {et~al.} 2015, \textit{A\&A}, 574, A14

\bibitem[Costa et al. (2014)]{Costa2014_z6_env_sims}
Costa, T., Sijacki, D., Trenti, M., \& Haehnelt, M.~G. 2014, \textit{MNRAS}, 439, 2146

\bibitem[De Rosa et al. (2011)]{DeRosa2011} 
De Rosa, G., Decarli, R., Walter, F., et al.\ 2011, \textit{ApJ}, 739, 56

\bibitem[De Rosa et al. (2014)]{DeRosa2014} 
De Rosa, G., Venemans, B.~P., Decarli, R., et al.\ 2014, \textit{ApJ}, 790, 145

\bibitem[Decarli et al. (2017)]{Decarli2017_z6_ALMA_comp} Decarli, R., Walter, F., Venemans, B.~P., et al.\ 2017, \textit{Nature}, 545, 457

\bibitem[Decarli et al. (2018)]{Decarli2018_CII_z6} 
Decarli, R., Walter, F., Venemans, B.~P., et al.\ 2018, \textit{ApJ}, 854, 97

\bibitem[Faisst et al. (2019)]{Faisst2019_ALPINE} 
Faisst, A.~L., Schaerer, D., Lemaux, B.~C., et al.\ 2019, arXiv e-prints, arXiv:1912.01621

\bibitem[Fan et al. (2003)]{Fan2003_z6} 
Fan, X., Strauss, M.~A., Schneider, D.~P., et al.\ 2003, \textit{AJ}, 125, 1649

\bibitem[Fan et al. (2006)]{Fan2006_reion_rev} 
Fan, X., Carilli, C.~L., \& Keating, B.\ 2006, \textit{ARA\&A}, 44, 415

\bibitem[Ghisellini et al. (2013)]{Ghisellini2013_blazars} 
Ghisellini, G., Haardt, F., Della Ceca, R., et al.\ 2013, \textit{MNRAS}, 432, 2818

\bibitem[Ghisellini et al. (2015)]{Ghisellini2015_blazar_z52} 
Ghisellini, G., Tagliaferri, G., Sbarrato, T., et al.\ 2015, \textit{MNRAS}, 450, L34

\bibitem[Giallongo et al. (2015)]{Giallongo2015_CANDELS_hiz_AGN} 
Giallongo, E., Grazian, A., Fiore, F., et al.\ 2015, \textit{A\&A}, 578, A83

\bibitem[Habouzit et al. (2019)]{Habouzit2019_z6_env} 
Habouzit, M., Volonteri, M., Somerville, R.~S., et al.\ 2019, \textit{MNRAS}, 489, 1206

\bibitem[Husband et al. (2013)]{Husband2013}
Husband, K., Bremer, M.~N., Stanway, E.~R., et al.\ 2013, \textit{MNRAS}, 432, 2869

\bibitem[Inayoshi et al. (2019)]{Inayoshi2020_seeds_rev} 
Inayoshi, K., Visbal, E., \& Haiman, Z.\ 2020, \textit{ARA\&A}, in press, arXiv:1911.05791

\bibitem[Izumi et al. (2018)]{Izumi2018_SHELLQs_III_z6_ALMA} 
Izumi, T., Onoue, M., Shirakata, H., et al.\ 2018, \textit{PASJ}, 70, 36

\bibitem[Izumi et al. (2019)]{Izumi2019_SHELLQs_VIII_z6_ALMA} 
Izumi, T., Onoue, M., Matsuoka, Y., et al.\ 2019, \textit{PASJ}, 71, 111

\bibitem[Janssen et al.(2016)]{Janssen2016_CII_outflows} 
Janssen, A.~W., Christopher, N., Sturm, E., et al.\ 2016, \textit{ApJ}, 822, 43


\bibitem[Jiang et al. (2006)]{Jiang2006_Spitzer} 
Jiang, L., Fan, X., Hines, D.~C., et al.\ 2006, \textit{AJ}, 132, 2127

\bibitem[Jiang et al. (2007)]{Jiang2007} 
Jiang, L., Fan, X., Vestergaard, M., et al.\ 2007, \textit{AJ}, 134, 1150

\bibitem[Jiang et al. (2010)]{Jiang2010_HDP} 
Jiang, L., Fan, X., Brandt, W.~N., et al.\ 2010, \textit{Nature}, 464, 380

\bibitem[Jiang et al. (2016)]{Jiang2016_SDSS_z6_final} 
Jiang, L., McGreer, I.~D., Fan, X., et al.\ 2016, \textit{ApJ}, 833, 222

\bibitem[Kim et al. (2009)]{Kim2009_idrops_z6} 
Kim, S., Stiavelli, M., Trenti, M., et al.\ 2009, \textit{ApJ}, 695, 809

\bibitem[Kim et al. (2018)]{Kim2018_z6_low_LLEdd} 
Kim, Y., Im, M., Jeon, Y., et al.\ 2018, \textit{ApJ}, 855, 138

\bibitem[Kormendy \& Ho (2013)]{KormendyHo2013_MM_Rev}
Kormendy, J., \& Ho, L.~C. 2013, \textit{ARA\&A}, 51, 511

\bibitem[Kulkarni et al. (2019)]{Kulkarni2019_QLF}
Kulkarni, G., Worseck, G., \& Hennawi, J.~F.\ 2019, \textit{MNRAS}, 488, 1035

\bibitem[Kurk et al. (2007)]{Kurk2007} 
Kurk, J.~D., Walter, F., Fan, X., et al.\ 2007, \textit{ApJ}, 669, 32

\bibitem[Leipski et al. (2014)]{Leipski2014_z6_FIR_SEDs} 
Leipski, C., Meisenheimer, K., Walter, F., et al.\ 2014, \textit{ApJ}, 785, 154

\bibitem[Luo et al. (2017)]{Luo2017_CDFS_7Ms} 
Luo, B., Brandt, W.~N., Xue, Y.~Q., et al.\ 2017, \textit{ApJS}, 228, 2

\bibitem[Lusso, \& Risaliti (2016)]{Lusso2016_Lx_Luv} 
Lusso, E., \& Risaliti, G.\ 2016, \textit{ApJ}, 819, 154

\bibitem[Maiolino et al. (2005)]{Maiolino2005_CII_J1148}
Maiolino, R., Cox, P., Caselli, P., et al.\ 2005, \textit{A\&A}, 440, L51

\bibitem[Maiolino et al. (2012)]{Maiolino2012_J1148_feedback} 
Maiolino, R., Gallerani, S., Neri, R., et al.\ 2012, \textit{MNRAS}, 425, L66


\bibitem[Marchesi et al. (2016)]{Marchesi2016_XVP_hiz} 
Marchesi, S., Civano, F., Salvato, M., et al.\ 2016, \textit{ApJ}, 827, 150

\bibitem[Marziani \& Sulentic (2012)]{MarzianiSulentic2012_MBH_rev} 
Marziani, P., \& Sulentic, J.~W.\ 2012, \textit{NewAR}, 56, 49

\bibitem[Mart{\'\i}nez-Aldama et al. (2018)]{MartinezAldama2018}
Mart{\'\i}nez-Aldama, M.~L., del Olmo, A., Marziani, P., et al.\ 2018, \textit{A\&A}, 618, A179


\bibitem[Matsuoka et al. (2019)]{Matsuoka2019_SHELLQs_X_z6} 
Matsuoka, Y., Iwasawa, K., Onoue, M., et al.\ 2019, \textit{ApJ}, 883, 183


\bibitem[Mazzucchelli et al. (2017)]{Mazzucchelli2017} 
Mazzucchelli, C., Ba{\~n}ados, E., Venemans, B.~P., et al.\ 2017, \textit{ApJ}, 849, 91

\bibitem[Mazzucchelli et al. (2019)]{Mazzucchelli2019_z6_comp} 
Mazzucchelli, C., Decarli, R., Farina, E.~P., et al.\ 2019, \textit{ApJ}, 881, 163

\bibitem[Miller et al. (2018)]{Miller2018_SPT_cluster_z43} 
Miller, T.~B., Chapman, S.~C., Aravena, M., et al.\ 2018, \textit{Nature}, 556, 469

\bibitem[Mor et al. (2012)]{Mor2012_z48} Mor, R., Netzer, H., Trakhtenbrot, B., et al.\ 2012, \textit{ApJL}, 749, L25

\bibitem[Nanni et al. (2017)]{Nanni2017} 
Nanni, R., Vignali, C., Gilli, R., et al.\ 2017, \textit{A\&A}, 603, A128

\bibitem[Natarajan (2011)]{Natarajan2011_seeds_rev} 
Natarajan, P.\ 2011, BASI, 39, 145

\bibitem[Natarajan et al. (2017)]{Natarajan2017_JWST_model} 
Natarajan, P., Pacucci, F., Ferrara, A., et al.\ 2017, \textit{ApJ}, 838, 117

\bibitem[Neeleman et al. (2019)]{Neeleman2019_z6_comp_ALMA} 
Neeleman, M., Ba{\~n}ados, E., Walter, F., et al.\ 2019, \textit{ApJ}, 882, 10

\bibitem[Netzer et al. (2014)]{Netzer2014_z48_SFR} Netzer, H., Mor, R., Trakhtenbrot, B., et al.\ 2014, \textit{ApJ}, 791, 34

\bibitem[Nguyen et al. (2020)]{Nguyen2020_z48_ALMA}
Nguyen, H.~N., Lira, P., Trakhtenbrot, B., et al.\ 2020, \textit{MNRAS}, submitted

\bibitem[Onoue et al. (2019)]{Onoue2019_SHELLQs_z6_MBH} 
Onoue, M., Kashikawa, N., Matsuoka, Y., et al.\ 2019, \textit{ApJ}, 880, 77

\bibitem[Overzier et al. (2006)]{Overzier2006_z52} 
Overzier, R.~A., Miley, G.~K., Bouwens, R.~J., et al.\ 2006, \textit{ApJ}, 637, 58

\bibitem[Overzier et al. (2009)]{Overzier2009} 
Overzier, R.~A., Guo, Q., Kauffmann, G., et al.\ 2009, \textit{MNRAS}, 394, 577

\bibitem[Pons et al. (2020)]{Pons2020_Xray_z65} 
Pons, E., McMahon, R.~G., Banerji, M., et al.\ 2020, \textit{MNRAS}, 491, 3884

\bibitem[Reed et al. (2017)]{Reed2017_z6_QSOs_DES_VISTA} 
Reed, S.~L., McMahon, R.~G., Martini, P., et al.\ 2017, \textit{MNRAS}, 468, 4702

\bibitem[Ross, \& Cross (2019)]{Ross2019_z6_phot_cat} 
Ross, N.~P., \& Cross, N.~J.~G.\ 2019, arXiv e-prints, arXiv:1906.06974

\bibitem[Sbarrato et al. (2012)]{Sbarrato2012_blazar_z53} 
Sbarrato, T., Ghisellini, G., Nardini, M., et al.\ 2012, \textit{MNRAS}, 426, L91

\bibitem[Shao et al. (2017)]{Shao2017_ALMA_z6} 
Shao, Y., Wang, R., Jones, G.~C., et al.\ 2017, \textit{ApJ}, 845, 138

\bibitem[Shemmer et al. (2006)]{Shemmer2006_CXO_hiz} 
Shemmer, O., Brandt, W.~N., Schneider, D.~P., et al.\ 2006, \textit{ApJ}, 644, 86

\bibitem[Shen(2013)]{Shen2013_rev} 
Shen, Y.\ 2013, \textit{BASI}, 41, 61

\bibitem[Shen et al. (2016)]{Shen2016_SDSS_RM_shifts} 
Shen, Y., Brandt, W.~N., Richards, G.~T., et al.\ 2016, \textit{ApJ}, 831, 7

\bibitem[Shen et al. (2019)]{Shen2019_z6_NIR_spec} 
Shen, Y., Wu, J., Jiang, L., et al.\ 2019, \textit{ApJ}, 873, 35

\bibitem[Shen et al. (2020)]{Shen2020_QLF} 
Shen, X., Hopkins, P.~F., Faucher-Gigu{\`e}re, C.-A., et al.\ 2020, arXiv e-prints, arXiv:2001.02696

\bibitem[Simpson et~al.(2014)]{Simpson2014_ULASJ1120_env}
Simpson, C., Mortlock, D., Warren, S., et~al.\ 2014, \textit{MNRAS}, 442, 3454

\bibitem[Stanley et al. (2019)]{Stanley2019_ALMA_z6_stack} 
Stanley, F., Jolly, J.~B., K{\"o}nig, S., et al.\ 2019, \textit{A\&A}, 631, A78

\bibitem[Stark (2016)]{Stark2016_hiz_gals_rev} 
Stark, D.~P.\ 2016, \textit{ARA\&A}, 54, 761

\bibitem[Stone et al.(2016)]{Stone2016_PACS_AGN_outflows} 
Stone, M., Veilleux, S., Mel{\'e}ndez, M., et al.\ 2016, \textit{ApJ}, 826, 111

\bibitem[Trakhtenbrot et al. (2011)]{Trakhtenbrot2011} 
Trakhtenbrot, B., Netzer, H., Lira, P., et al.\ 2011, \textit{ApJ}, 730, 7

\bibitem[Trakhtenbrot \& Netzer (2012)]{TrakhtNetzer2012_Mg2} 
Trakhtenbrot, B., \& Netzer, H.\ 2012, \textit{MNRAS}, 427, 3081

\bibitem[Trakhtenbrot et al. (2016)]{Trakhtenbrot2016_COSMOSFIRE_MBH} 
Trakhtenbrot, B., Civano, F., Urry, C.~M., et al.\ 2016, \textit{ApJ}, 825, 4

\bibitem[Trakhtenbrot et al. (2017a)]{Trakhtenbrot2017_z6_Mdot_eta}
Trakhtenbrot, B., Volonteri, M., \& Natarajan, P.\ 2017a, \textit{ApJL}, 836, L1

\bibitem[Trakhtenbrot et al. (2017b)]{Trakhtenbrot2017_z48_ALMA} 
Trakhtenbrot, B., Lira, P., Netzer, H., et al.\ 2017b, \textit{ApJ}, 836, 8


\bibitem[Utsumi et~al.(2010)]{Utsumi2010_env_J2329}
Utsumi, Y., Goto, T., Kashikawa, N., et~al.\ 2010, \textit{ApJ}, 721, 1680

\bibitem[Valiante et al. (2017)]{Valiante2017_seeds_rev} 
Valiante, R., Agarwal, B., Habouzit, M., et al.\ 2017, \textit{PASA}, 34, e031

\bibitem[Veilleux et al. (2013)]{Veilleux2013_Herschel_outflows} 
Veilleux, S., Mel{\'e}ndez, M., Sturm, E., et al.\ 2013, \textit{ApJ}, 776, 27

\bibitem[Venemans et al. (2012)]{Venemans2012_z71_CII} 
Venemans, B.~P., McMahon, R.~G., Walter, F., et al.\ 2012, \textit{ApJL}, 751, L25

\bibitem[Venemans et al. (2015)]{Venemans2015_PS1_z6} 
Venemans, B.~P., Ba{\~n}ados, E., Decarli, R., et al.\ 2015, \textit{ApJL}, 801, L11

\bibitem[Venemans et al. (2016)]{Venemans2016_z6_cii}
Venemans, B.~P., Walter, F., Zschaechner, L., et al.\ 2016, \textit{ApJ}, 816, 37

\bibitem[Venemans et al. (2017)]{Venemans2017_z75_CII} 
Venemans, B.~P., Walter, F., Decarli, R., et al.\ 2017, \textit{ApJL}, 851, L8

\bibitem[Venemans et al.(2018)]{Venemans2018_z6_ALMA_cont} 
Venemans, B.~P., Decarli, R., Walter, F., et al.\ 2018, \textit{ApJ}, 866, 159

\bibitem[Venemans et al. (2019)]{Venemans2019_z66_ALMA_hires} 
Venemans, B.~P., Neeleman, M., Walter, F., et al.\ 2019, \textit{ApJL}, 874, L30


\bibitem[Vito et al. (2016)]{Vito2016_CDFS_7Ms} 
Vito, F., Gilli, R., Vignali, C., et al.\ 2016, \textit{MNRAS}, 463, 348

\bibitem[Vito et al. (2018)]{Vito2018_hiz_XLF_fobs} 
Vito, F., Brandt, W.~N., Yang, G., et al.\ 2018, \textit{MNRAS}, 473, 2378

\bibitem[Vito et al. (2019)]{Vito2019_z6_Xrays} 
Vito, F., Brandt, W.~N., Bauer, F.~E., et al.\ 2019, \textit{A\&A}, 630, A118

\bibitem[Volonteri (2010)]{Volonteri2010_rev}
Volonteri, M.\ 2010, \textit{A\&ARv}, 18, 279

\bibitem[Volonteri et al. (2011)]{Volonteri2011_blazars} 
Volonteri, M., Haardt, F., Ghisellini, G., et al.\ 2011, \textit{MNRAS}, 416, 216

\bibitem[Volonteri et al. (2015)]{Volonteri2015_superEdd} 
Volonteri, M., Silk, J., \& Dubus, G.\ 2015, \textit{ApJ}, 804, 148

\bibitem[Volonteri et al. (2017)]{Volonteri2017_JWST_model} 
Volonteri, M., Reines, A.~E., Atek, H., et al.\ 2017, \textit{ApJ}, 849, 155

\bibitem[Walter et al.(2018)]{Walter2018_OIII_88mic_z6} 
Walter, F., Riechers, D., Novak, M., et al.\ 2018, \textit{ApJL}, 869, L22

\bibitem[Wang et al. (2015)]{Wang2015_z5_hiM} 
Wang, F., Wu, X.-B., Fan, X., et al.\ 2015, \textit{ApJL}, 807, L9

\bibitem[Wang et al. (2016)]{Wang2016_WISE_z5_qso} 
Wang, F., Wu, X.-B., Fan, X., et al.\ 2016, \textit{ApJ}, 819, 24

\bibitem[Wang et al. (2013)]{Wang2013_z6_ALMA} 
Wang, R., Wagg, J., Carilli, C.~L., et al.\ 2013, \textit{ApJ}, 773, 44

\bibitem[Wang et al. (2019a)]{Wang2019_z65_QLF} 
Wang, F., Yang, J., Fan, X., et al.\ 2019, \textit{ApJ}, 884, 30

\bibitem[Wang et al. (2019b)]{Wang2019_z6_ALMA_hires} 
Wang, R., Shao, Y., Carilli, C.~L., et al.\ 2019, \textit{ApJ}, 887, 40

\bibitem[Weigel et al. (2015)]{Weigel2015_CDFS} 
Weigel, A.~K., Schawinski, K., Treister, E., et al.\ 2015, \textit{MNRAS}, 448, 3167

\bibitem[Willott et al. (2005)]{Willott2005_z6_comp} 
Willott, C.~J., Percival, W.~J., McLure, R.~J., et al.\ 2005, \textit{ApJ}, 626, 657

\bibitem[Willott et al. (2010a)]{Willott2010_QLF} 
Willott, C.~J., Delorme, P., Reyl{\'e}, C., et al.\ 2010a, \textit{AJ}, 139, 906

\bibitem[Willott et al. (2010b)]{Willott2010_MBH} 
Willott, C.~J., Albert, L., Arzoumanian, D., et al.\ 2010b, \textit{AJ}, 140, 546

\bibitem[Willott et al. (2013)]{Willott2013_z6_ALMA} 
Willott, C.~J., Omont, A., \& Bergeron, J.\ 2013, \textit{ApJ}, 770, 13

\bibitem[Willott et al. (2015)]{Willott2015_CFHQS_ALMA} 
Willott, C.~J., Bergeron, J., \& Omont, A.\ 2015, \textit{ApJ}, 801, 123

\bibitem[Willott et al. (2017)]{Willott2017_z6_SFR} 
Willott, C.~J., Bergeron, J., \& Omont, A.\ 2017, \textit{ApJ}, 850, 108

\bibitem[Wu et al. (2015)]{Wu2015_z6_nature} 
Wu, X.-B., Wang, F., Fan, X., et al.\ 2015, \textit{Nature}, 518, 512

\bibitem[Yang et al.(2018)]{Yang2018_z55_survey} 
Yang, J., Wang, F., Fan, X., et al.\ 2018, arXiv e-prints, arXiv:1810.11927








\end{thebibliography}

\bibliographystyle{mn2e}







\end{document}